# Quantifying the Risk of Pastoral Conflict in 4 Central African Countries

Lirika Sola[a], Youdinghuan Chen[a], Samantha K. Murphy [b,1], V.S. Subrahmanian[a]*

*a Northwestern University, Department of Computer Science and Buffett Institute for Global Affairs, Evanston, IL, USA*

*b Cambridge University, Department of Geography, Cambridge, United Kingdom*

\* vss@northwestern.edu

---

[1] *Work conducted while at United Nations Department of Political and Peace-Building Affairs, New York, NY, USA*

# Quantifying the Risk of Pastoral Conflict in 4 Central African Countries

**Abstract**

Climate change is becoming a widely recognized risk factor of farmer-herder conflict in Africa. Using an 8-year dataset (Jan 2015–Sep 2022) of detailed weather and terrain data across four African nations, we apply statistical and machine learning methods to analyze pastoral conflict. We test hypotheses linking these variables with pastoral conflict within each country using geospatial and statistical analysis. Complementing this analysis are risk maps automatically updated for decision-makers. Our models estimate which cells have a high likelihood of experiencing pastoral conflict with high predictive accuracy and study the variation of this accuracy with the granularity of the cells.

Word Count: abstract (100), total (12635)

**Introduction**

Pastoralism is a traditional practice that existed since time immemorial. Depictions of herders and their livestock can be seen in cave art from over 5000 years ago.[1] Shepherding is referenced in an array of ancient works from ancient Mesopotamian, Achaemenid, and Greek cultures[2] as well as in the Old Testament.[3] "Transhumans" or "pastoralists" have become contemporary terms for these nomadic herders who wander ancient trails in search of food and water for their flock.[4]

The rising threat of climate change in the last few decades is changing the availability of vegetation and water for livestock, altering routes pastoralists have travelled for, in some cases, hundreds to thousands of years.[5] As a consequence, pastoralists (who are nomadic) and farmers (who are sedentary), are struggling to maintain the once harmonious relations they established through historic trade, resource coordination, and informal land access agreements. As pastoralists seek out vegetation and water for their livestock, they are often seen as an encroaching onto sedentary communities, driving protectionism and even conflict over resources.[6]

In this study we examine the risk of pastoral conflicts using geospatial analysis in conjunction with detailed terrain and meteorological data from NASA which is far more fine-grained than past work discussed in the "Pastoral Conflicts in Central Africa: A Brief History" section. Additionally, we utilize historical conflict data over a 93-month time horizon (Jan 2015 to Sep 2022). Our study differs from past work in the following respects:

- We are the first to bring advanced machine learning techniques to understand and characterize the risk of pastoral conflict in cells at different granularities (50km x 50km, 75km x 75km, 100km x 100km). We are able to predict pastoral conflict risk for Cameroon with high success at the (100km x 100km) granularity with an F1-score of 91% and an AUC of 99%, as well as for the DRC (F1-Score 77%, AUC 98%). We can make highly accurate predictions at the (75km x 75km) granularity for CAR (F1-Score 72%, AUC 95%) and a bit less well for Chad (F1-score 63%, AUC 88%). The trade-off between granularity of cells and the quality of prediction is, unsurprisingly, not a clean one.

- We are the first to use dependent variables that focus on pastoral conflict[2] in a statistical and machine learning study.

- We are the first to curate fine-grained terrain and meteorological independent variables using NASA data as compared to coarse-grained metrics used by others. **Our data, called *PCORE* (short for Pastoral Conflict Reasoning Engine) will be publicly released when this paper is published**.

- We propose and examine both univariate and complex multivariate hypothesis and learn indicators of pastoral conflict at a country-level. In our statistical analysis, we perform stratification on the specific country/region. This approach provides a stratum-specific risk estimate with respect to conflict, which offers nuanced and more precise inference about each underlying country as opposed to a general approach that covers lots of countries.

- Our study is rooted in 4 Central African countries over a period of 93 months (Jan 2015 to Sep 2022): Cameroon, Central African Republic (CAR), Chad, and the Democratic Republic of Congo (DRC) and report specific results for these countries for the first time.

---

[2] In contrast, some past studies model *all* conflicts in areas where pastoral activity overlaps farmer activity which potentially introduces other kinds of conflicts into the study, not just pastoral ones.

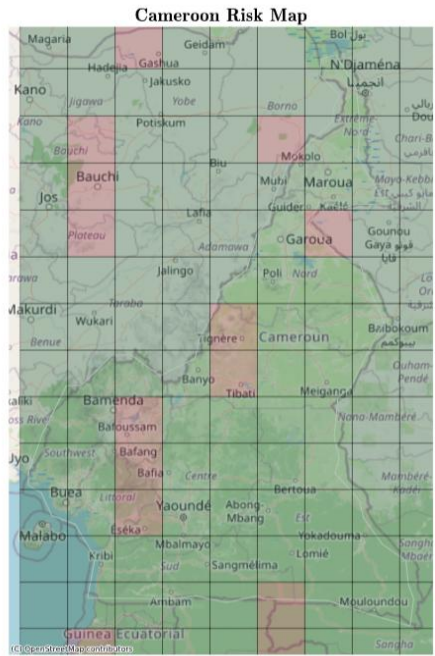
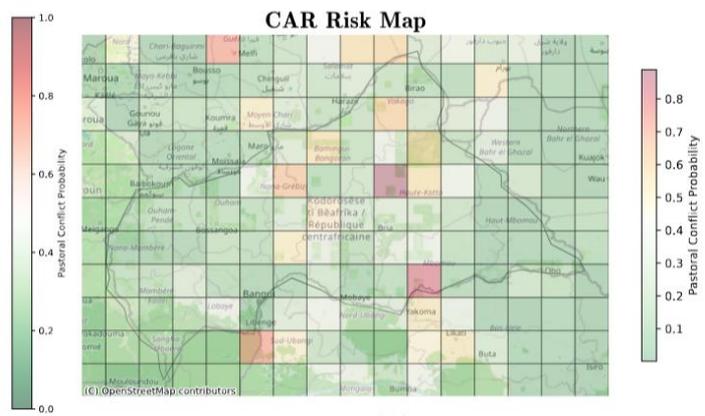
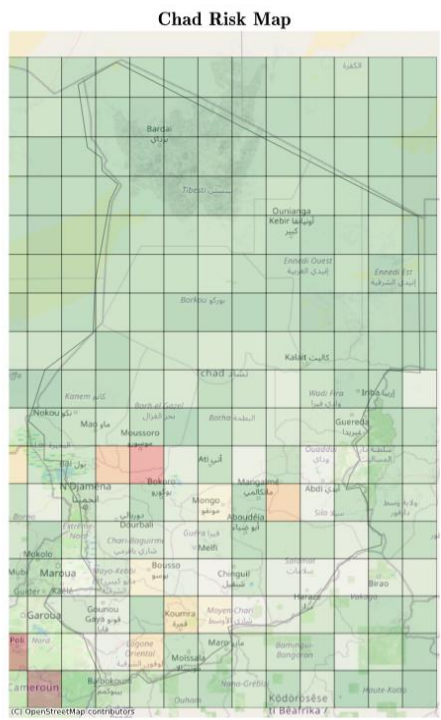
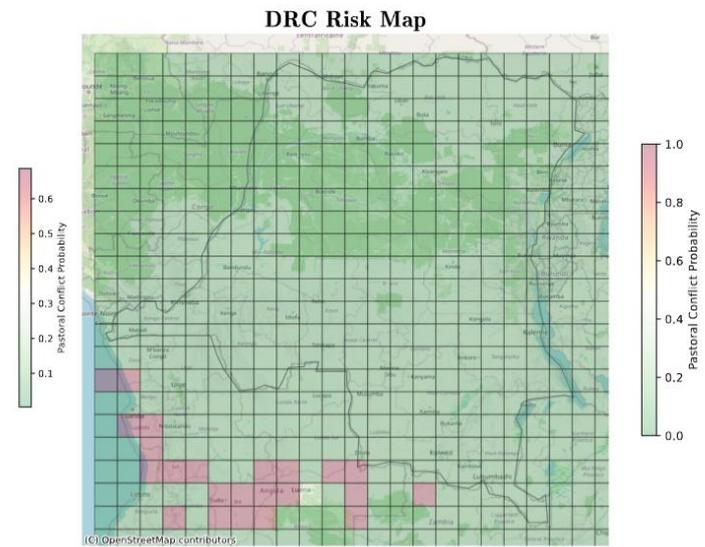

**Figure 1**. Map showing the Risk of a (100 × 100) km cell experiencing a pastoral conflict in (a) Cameroon, (b) CAR, (c) Chad and (d) DRC. The color of a cell represents the likelihood of a pastoral conflict occurring.

| Variable | Cameroon | CAR | Chad | DRC |
|---|---|---|---|---|
| Maximum Wind Speed at 10m | ✗ | ✓★★ | ✓★ | ✓★★ |
| Minimum Wind Speed at 10m | ✗ | ✓★★ | ✓★★ | ✓★★★ |
| Relative Humidity at 2m | ✗ | ✓★★ | ✓★★ | ✓★★ |
| Corrected Precipitation | ✗ | ✗ | ✓★★★ | ✗ |
| Maximum Temperature at 2m | ✗ | ✓★★ | ✓★★★ | ✓★ |
| Minimum Temperature at 2m | ✗ | ✓★★ | ✓★★★ | ✓★ |
| Leaf Area Index | ✓★★ | ✓★★ | ✓★★★ | ✓★★ |
| Greenness | ✗ | ✓★★ | ✓★★ | ✓★★ |
| Surface Soil Wetness | ✗ | ✓★★★ | ✓★★ | ✗ |
| Surface Skin Temperature | ✗ | ✓★★★ | ✓★★ | ✓★★★ |
| Land Evaporation | ✓★★ | ✓★★★ | ✓★★★ | ✓★ |

**Table 1.** Relationship between terrain and meteorological variables and the occurrence of pastoral conflicts.

Table 1 summarizes the results linking key variables and their relationship to pastoral conflict that passed statistical muster. We now discuss these below.

(1) **The role of weather**: We considered three broad sets of independent variables in our study: temperature, humidity, and wind conditions.

   a. The role of *temperature* in assessing risk of pastoral conflict in each zone is highly nuanced. In Cameroon, we find no link between temperature and pastoral conflict, while in CAR, Chad and DRC, we find that both lower and higher temperatures are linked to higher probability of pastoral conflict with temperatures "in the middle" not linked to pastoral conflict.

b. The role of *humidity* is likewise highly nuanced. In Cameroon we find no link between humidity and pastoral conflict, while in CAR, Chad and the DRC, we find that both low humidity and high humidity are linked to pastoral conflict with mid-range humidity not being linked to pastoral conflict.

c. The role of *wind conditions* in assessing pastoral conflict risk is also nuanced. In Cameroon we again find no link whatsoever. But in CAR, Chad and the DRC, we find that both low wind and high wind are linked to pastoral conflict with mid-range wind conditions not being linked to pastoral conflict.

(2) **The role of terrain**: We considered three broad sets of variables in our study: vegetation, soil wetness, evaporation, and temperature.

   a. The role of *vegetation* in assessing risk of pastoral conflict is measured through the Leaf Area Indexed (LAI) and Greenness Fraction (GRN) metrics. In Cameroon, areas with lower amounts of vegetation are more likely to experience pastoral conflict. In CAR, the Leaf Area Indexed indicates that both lower and higher values of it are connected to higher probability of pastoral conflicts occurring. On the other hand, in Chad both lower and higher values of the Greenness Fraction and Leaf Area Indexed are connected to higher risks of pastoral conflicts. In the DRC, areas with both low and high vegetation density are more likely to experience pastoral conflict according to the LAI metric and Greenness Fraction.

   b. The role of *soil wetness* in assessing risk of pastoral conflict is measured through the Surface Soil Wetness (SSW) metric and the Land

Evaporation (LNDEV) metrics. In Cameroon and the DRC, less wet regions are more likely to experience pastoral conflict, while in CAR and Chad, both lower and higher rates of land evaporation are linked to higher probability of pastoral conflicts. As for the Surface Soil Wetness, in Cameroon and the DRC there is no relation between pastoral conflicts and the soil wetness. In CAR, lower values of soil wetness are related to higher likelihood of pastoral conflicts while in Chad, the opposite occurs.

    c. The role of *Surface Skin Temperature* (SST) in assessing risk of pastoral conflict is also complex. It plays no role in Cameroon, but when it is very low, it is positively linked to pastoral conflict in CAR and DRC. In Chad both lower and higher values of the variable are linked to higher likelihood of pastoral conflicts occurring.

    d. The role of *Land Evaporation* (LNDEV) is also nuanced. In Cameroon and DRC, lower values of land evaporation are linked to a higher probability of pastoral conflicts occurring. In CAR and Chad, both lower and higher values of land evaporation rate are linked to an increased risk of pastoral conflicts taking place.

## *Pastoral Conflicts in Central Africa: A Brief History*

Extending from the arid Sahel to the lush wetlands of the Congo Basin, the eleven-country region of Central Africa lies at heart of pan-African transhumance. The ecologically diverse region sits at the frontier of desert and pasture, which pastoralists have cyclically migrated between for centuries. Moving in harmony with the wet and dry seasons, pastoralists navigate across national and subnational borders to nourish their livestock, engage in trade, and keep their cultural heritage alive. This tradition is as much an economic activity as it is intrinsically linked to indigenous identity. At its core, transhumance is a form of environmental adaptation whereby mobility is responsive to naturally occurring drought, rainfall patterns, and other environmental conditions.

Although transhumant pastoralism in Central Africa can be traced back to the mid-18th century, it likely extends much further into Africa's historical record. For possibly millennia, pastoralists have developed routes linking West Africa, North Africa, and East Africa. These migratory corridors are still used today, albeit with increased risks: from the eastern countries of Sudan and Uganda pastoralists migrate southwest through high-altitude wetlands to the Democratic Republic of the Congo (DRC) meanwhile pastoralists from Niger and northern Nigeria migrate east to the Logone floodplain during the dry seasons. Sudano-Sahelian pastoralists travel within and beyond the borders of the Central African Republic (CAR). Pastoralism is a historically important economic, political, and social activity for Central African nations, especially for those at the centre of regional interaction zones which we study in this paper: Chad, CAR, Cameroon, and the DRC. [7, 8, 9]

The livestock industry contributes significantly to the region's economy. In the DRC, cattle farming represents 11% of the country's potential livestock capacity of 40 million animals and contributes 9.2% to the agricultural GDP.[13] In Cameroon, the sector accounts for 13% of the national GDP and employs over 30% of its rural population.[10] In CAR the sector contributes as much as 15% to the national GDP.[14] Livestock production accounts for approximately 25% of Chad's GDP.[11, 12]

These economic contributions are supported by pastoral systems that manage 80% of the region's livestock using a mobile model optimized for the seasonal availability of pastures. These systems involve regular migrations of herds between fixed points such as seasonal grazing areas, water sources, villages, and markets. Transhumance typically commences just before the dry season, moving from north to south, and is strategically planned to optimize routes to the destination areas.[7] The return north begins before the rainy season, adjusted according to obstacles like river crossings, weather patterns, and harvest schedules.

While these migrations have been historically responsive to changing environmental conditions, it is becoming harder to adapt to shifting weather patterns and rising political uncertainty. More frequent natural disasters, border closures, and land access issues are just a few factors that distort pastoral migration. Prolonged drought and resource scarcity driven by climate change are pushing herdsmen into different territories, creating new contact zones with hostile actors, whether sedentary communities on privatized land, non-state armed groups, or along unstable border areas rife with conflict. Even in relatively stable areas, the presence of pastoralists can be seen as encroachment and heighten the risk of emergent conflicts driven by competition over scarce natural resources.[15, 16, 17, 18]

Shifting climate patterns are limiting access to critical resources like water and arable land, putting pressure on pastoral and agricultural communities alike. Water sources have shrunk drastically across Central Africa since the 1960s because of rising temperatures, heightened irrigation demand, and erratic rainfall. One of the region's most important agroecological zones supporting the food security of millions of people, Lake Chad, has become a hotspot of climate insecurity and conflict after losing approximately 90% of its area because of human and climatic factors.[19, 20, 54, 12] Although the basin has shown signs of replenishment over the past five years, the long term security and livelihood impacts of volatility are not without consequence; intercommunal friction rooted in competition and the perception of scarcity continues to undermine stability. As a consequence, scarcity heightens competition among local communities, particularly in pastoral areas where water is vital for human and livestock sustenance and deeply ingrained in their livelihoods and cultural practices.[54, 54]

Besides water scarcity, rising temperatures are becoming a greater risk factor of regional instability.[24] Higher temperatures accelerate desertification, particularly in the semi-arid regions of Central Africa, transforming once fertile lands into deserts. This loss of arable land limits agricultural productivity, exacerbating food and economic insecurity. Livestock suffer from heat stress, reducing their fertility and milk production, undermining the economic foundation of pastoral communities. Increased heat also raises the overall demand for water while accelerating evapotranspiration, intensifying the strain on already scarce resources. As water supply diminishes and demand rises, agency at the individual and communal levels is diminished, potentially heightening preexisting socioeconomic tensions and propensity for conflict.[25]

*Cameroon*: In Cameroon, farmer-herder conflicts are rooted in political and environmental challenges. They are linked to disputed and constantly changing legal structures governing access to land and the right of movement, often associated with corrupt and biased postcolonial power dynamics.[26] Environmental factors like land degradation, climate change, and population growth further exacerbate these tensions.[26, 27] Rainy seasons became shorter and more erratic as a result of climate shifts, leading to reduced grassland availability and intensified competition over scarce land and water resources.[28] This has led to violent clashes, including the 1981 Wum police brutality case and the 2017 attacks on Fulbe pastoralists by armed Mambila farmer groups on the Nigeria-Cameroon border, resulting in significant casualties and displacement.[26]

*CAR*: Although CAR is predicted to have relatively mild physical effects from climate change when compared to the rest of the continent, the combination of rapid population growth, insufficient adaptation capacities, and ongoing conflicts worsens the impact.[29] The country is expected to see more extreme temperatures and heatwaves, notably in the north and central regions, as well as greater rainfall variability, leading to more frequent floods and droughts concentrated in the central and southern parts.[30] Extreme weather events, such as the torrential rains in Bangui in July 2022, which displaced more than 21,000 people, are leading to increased internal displacement.[31] Furthermore, changes in traditional pastoral mobility, particularly the southward expansion of transhumance from the Sahel, are increasing competition for land, especially as herders come during harvest season.[32] In CAR, inadequate government control and the existence of mineral resources have transformed the northwest and eastern regions into primary location zones for armed groups, weakening traditional conflict resolution and exacerbating farmer-herder conflicts.[33,29]

*Chad*: Climate change is intensifying transhumance-related conflicts in Chad through environmental shifts such as the southward expansion of the Sahara, rising temperatures, and more erratic rainfall, which disrupt traditional herding routes and schedules.[34] Herders are moving south earlier, often before crops are harvested, and staying longer in regions densely populated by farmers.[34] This has led to increased tensions and violent encounters, including major incidents in 2019 and 2020. Over a hundred lives were lost in just nine days in August 2019, and subsequent conflicts in Salamat and Kabbia provinces of Chad resulted in dozens more casualties, prompting a state of emergency and border closures to manage the escalating situation.[35]

*DRC*: Despite its wealth of natural resources, the DRC remains one of the world's poorest countries, heavily affected by climate challenges such as heat waves, land degradation, and erratic rainfall. Rising temperatures are undermining food security, as drought and extreme heat degrade land,[36] reduce agricultural yields, and deplete natural resources, forcing many indigenous groups into a more nomadic lifestyle. This increased mobility exacerbates land disputes with other communities and complicates land recognition processes. Although the communities have traditionally practiced semi-nomadism, the growing need for greater mobility has escalated pastoral conflicts over land and resources.[37, 38]

*Other African Countries*: Transhumant pastoralism in both East and West Africa is severely impacted by climate-induced resource scarcity, which has intensified conflicts among communities. For example, in Northern Kenya, continuous climate change has led to prolonged droughts that drastically altered traditional pastoral patterns. The regional communities are forced to migrate more frequently to search for water and pasture that were already scarce. The process is highly competitive and often led to conflicts involving violence and cattle rustling with larger-scale socioeconomic disruptions.[39, 40] Similarly, in Southern Ethiopia's Borana area and Uganda's Karamoja region, conditions of severe drought diminished the natural resources essential for sustainable pastoralism.[36, 41]

In West African countries such as Ghana and Nigeria, the conflicts between farmers and herders (e.g. the Fulani people) have been well-studied due to intensive research efforts in characterizing resource scarcity and climate change.[42, 43] These conflicts often arise from disputes over farming and grazing lands. In Côte d'Ivoire (Ivory Coast) and Ghana, hired Fulani herders are frequently accused of cattle theft.[44, 45] In Nigeria and Kenya,[46, 47, 48] common issues include crop destruction by cattle, cattle rustling, and farmers encroaching on established grazing routes and lands. These incidents are further complicated by additional underlying factors, such as climate change and arms smuggling across porous borders, which intensify the severity and frequency of these conflicts. These clashes are often at a considerable scale with fatalities comparable to those from acts of terror by terrorist groups such as Al-Qaeda,

Al Shabab, and Boko Haram.[8] Taken together, pastoral conflicts impose a large impact on the environment, stability and socio-economic states of the region and deserve in-depth investigation.

The significance of pastoralism to the economies of Central African countries cannot be overstated, particularly in light of how changing climate patterns have diminished essential resources, critically affecting the livelihoods of pastoral communities. This paper seeks to provide a data-driven, quantitative analysis of the risk of pastoral conflicts in 4 Central African nations (Cameroon, CAR, Chad, and DRC) using a combination of statistical and machine-learning techniques.

*Methods*

*Our Data, Jan 2015 to Sep 2022*

We created the (Pastoral Conflict Reasoning Engine), a comprehensive dataset for each of the four countries in our study, covering the period from January 2015 to September 2022. For our statistical analysis, we divided each country into a grid of ($100km \times 100km$) cells to uniformly analyse the impact of various factors. For our machine learning analysis, we divided countries at 3-levels of granularity: ($100km \times 100km$) cells, ($75km \times 75km$) cells, and ($50km \times 50km$) cells. Each dataset aggregates diverse open-source information for each grid cell, including terrain characteristics, meteorological conditions, and historical pastoral conflict data from neighboring areas.

*Pastoral Conflicts, 2015-2022*

We created our pastoral conflict (dependent variable) data from the Armed Conflict Location & Event Data Project (ACLED).[49] Our dataset documents a total of 630 conflicts across Chad, the DRC, Cameroon, and CAR. Figure 2(a) displays the total number of pastoral conflicts in each country, with the DRC experiencing the most (280 pastoral conflicts), with Chad coming in second with 173. The selection process for identifying pastoral conflicts within the ACLED dataset involved filtering records based on the 'country' column and manually reviewing the 'notes' column, which contains detailed descriptions of the events, to ensure accurate categorization. In some cases, we also consulted other sources (e.g. reputable news sources through Nexis Uni) when curating the data.

Figure 2(b) categorizes the pastoral conflict data by year and country, highlighting significant fluctuations in conflict frequency and notable increases in recent years. During the 2015-2019 period, pastoral conflict counts remained relatively low and stable across all four countries. In Chad, pastoral conflicts were sporadic with slight increases in 2016 and 2017. Cameroon and CAR exhibited consistent pastoral conflict activity, with CAR experiencing slightly higher levels than other countries. The DRC had the lowest pastoral conflict counts during this period, with a gradual increase toward 2019.

In 2020, a dramatic increase in pastoral conflict counts was observed across all countries, particularly in Chad and the DRC, with the latter exceeding 120 conflicts, marking the highest in the dataset.

From 2021 to 2022, despite a slight decrease from the 2020 peak, pastoral conflict counts remained high in CAR, Chad, and the DRC, indicating a period of sustained instability. Cameroon also experience high pastoral conflict counts through these years, suggesting ongoing issues.

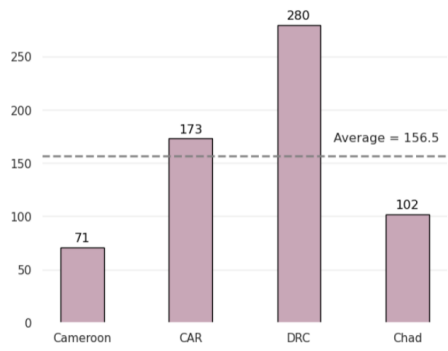
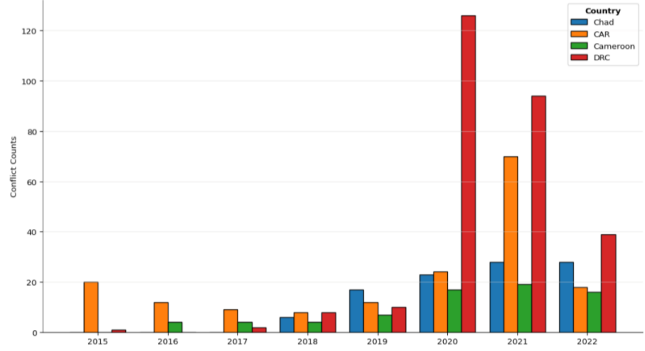

**Figure 2.** Number of Pastoral Conflicts, 2015 - 2022. (a) shows the total number of pastoral conflicts per country. (b) shows the yearly number of pastoral conflicts per each country. 2022 numbers only run through the end of September 2022. These charts are based on ACLED data. [49]

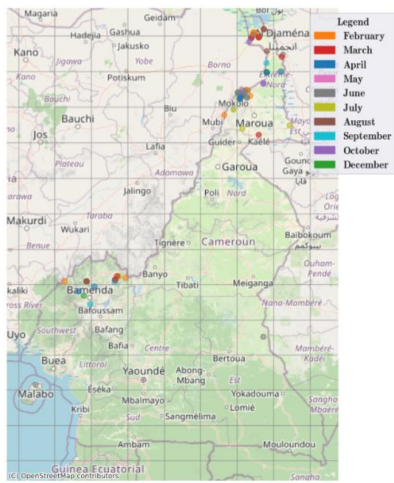
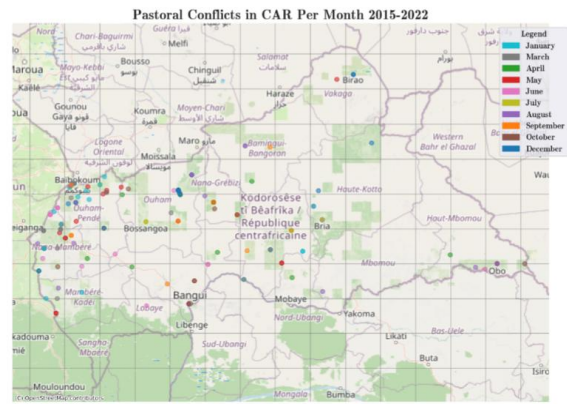
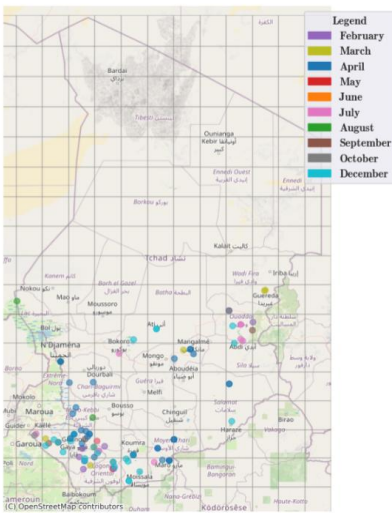
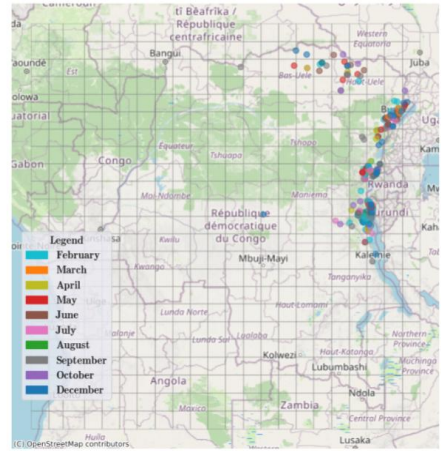

**Figure 3.** Month-by-Month Locations of Pastoral Conflicts, January 2015 - September 2022. The color of a dot shows the month when a pastoral conflict occurred. These charts are based on ACLED data. [49]

*Cameroon*: Figure 3 shows that the northern areas of Cameroon, especially around cities like Maroua and Garoua, have a consistent high level of conflicts throughout the year. Moreover, significant conflict activity is also observed in cities like Bamenda in the Western region and Bafoussam in the Central region, indicating that pastoral conflicts have extended from the typically dry northern regions to more central and western areas of the country. The northern regions, characterized by a semi-arid climate and part of the Sahel, face significant pressure due to strong climatic variations and irregular rainfall. These conditions lead to severe droughts and environmental crises, which heavily impact pastoral activities.[50] This suggests that conflicts tend to increase during the dry season, from November to April, when the competition for scarce resources such as water and grazing land intensifies, leading to heightened tensions across the regions.

*CAR*: Figure 3 shows that in CAR, there is a higher concentration of pastoral conflicts in the northwestern regions, such as Ouham and Nana-Grébizi. These areas are historically prone to conflicts due to ethnic tensions and competition over grazing lands.[51] Our data visualization suggests no clear seasonal pattern but does show that January, May, June and December experience more pastoral conflicts. This could be influenced by migration patterns and seasonal changes in pasture availability.

*Chad*: Figure 3 shows that the Central and Eastern parts of Chad, particularly around cities like Am Timan and Mongo, experience many pastoral conflicts. This region borders Sudan and is influenced by cross-border dynamics including migration and cattle raiding. Pastoral conflicts are consistently reported throughout the year with some areas showing prolonged conflict periods. This could suggest underlying issues such as land disputes, ethnic tensions, or political instabilities that exacerbate during certain periods.[52]

*DRC*: Figure 3 shows that in the DRC, pastoral conflicts are concentrated in the eastern part, particularly along the borders with Uganda, Rwanda, and Burundi. Notable clusters of pastoral conflicts are found in regions such as Haut-Uele, North Kivu, and South Kivu. There is a noticeable increase in pastoral conflicts during the middle of the year, with May, June, and July (represented by red, pink, and purple respectively) showing a higher frequency of incidents. This suggests a seasonal peak in pastoral conflicts during these months. Additionally, July and August (represented by purple and green) also exhibit a significant number of pastoral conflicts, further reinforcing the mid-year peak.

*Weather Data*

For each grid cell across the countries studied, we collected several weather-related factors using the NASA POWER Project (see https://power.larc.nasa.gov/). The weather factors considered include:

- Maximum Wind Speed at 10 Meters (WS10M_MAX): The maximum wind speed at 10 meters above the earth's surface. [53]

- Minimum Wind Speed at 10 Meters (WS10M_MIN): The minimum wind speed at 10 meters above the earth's surface. [53]

- Relative Humidity at 2 Meters (RH2M): The ratio of the actual partial pressure of water vapor to the saturation partial pressure, expressed as a percentage. [53]

- Corrected Total Precipitation (PRECTOTCORR): The bias-corrected total sum of precipitation at the earth's surface. [53]

- Maximum Temperature at 2 Meters (T2M_MAX): The maximum temperature of the air (dry bulb) at 2 meters above the earth's surface. [53]

- Minimum Temperature at 2 Meters (T2M_MIN): The minimum temperature of the air (dry bulb) at 2 meters above the earth's surface. [53]

*Histogram-based Features*

For each variable $v$, we split the interval between the minimum value of the variable and the maximum value in the dataset into 10 equal-sized parts. We then represented the variable via a histogram capturing the distribution of the variable's values over time. This format prepares the data for use in our machine learning and statistical analyses, allowing us to effectively assess the effects of weather variations on the occurrence of pastoral conflicts.

*Terrain Data*

In a similar manner to the weather-related features, an array of terrain-related features was gathered monthly for each cell within the countries' grids using NASA GES DISC (see https://disc.gsfc.nasa.gov/) data. The same histogram representation approach was employed to standardize this data, effectively capturing the variability of terrain conditions in a format that is compatible with our machine learning models. The terrain features considered include:

- Leaf Area Index (LAI): Measures the total leaf area relative to the ground area. The leaf area index is a key indicator of plant growth, impacting pastoral activities and movements. It is presented in a unit-less format. [54]

- Greenness Fraction (GRN): Represents the percentage of ground covered by living vegetation. This metric is crucial for evaluating the coverage and health of vegetation, making it vital for agricultural monitoring at both regional and global scales. [55]

- Surface Soil Wetness (SSW): Given in percentage saturation, this variable is critical for controlling the exchange of water and heat with the atmosphere, influencing both temperature and humidity. It supports plant health and helps

understand broader environmental impacts such as weather patterns, droughts, and flood susceptibility. [56]

- Surface Soil Temperature (SST): This quantifies the thermal energy emitted from the earth's surface, which is influenced by solar radiation. It can be used to identify different land cover types, monitor environmental changes, and assess the effects of land changes on global climate phenomena, including heatwaves and droughts in sensitive regions. [57]

- Land Evaporation (LNDEV): Significantly affects climate studies, water management, and weather prediction models. It is impacted by several factors, such as rainfall, atmospheric temperature, solar energy, and vegetation density. Variations in crop characteristics and soil management practices also markedly influence evaporation rates. [58]

*Nearby Pastoral Features*

To enrich our dataset, we introduced additional variables that reflect the spatial distribution of conflicts relative to each grid cell.

Let Nbr($j, c$) denote the set of all cells within a Euclidean distance j from cell c, excluding c itself. For example, for $j = 2$ and $c = (20,20)$, Nbr($j, c$) includes the cells (18,20), (18,19), (19,20), (20,21), (20,22), (21,21) but not (21,22). For each distance $j = 1, \ldots, 5$, a binary variable indicates whether any cell within Nbr($j, c$) experienced a conflict during the training period. Additionally, a numerical variable quantifies the total number of pastoral conflicts occurring within these neighboring cells. These variables are designed to capture both the presence and intensity of pastoral conflicts around each cell, thus providing a more detailed perspective on the spatial dynamics of this type of conflict.

*Univariate Relationships Between Weather and Terrain Features vs. Pastoral Conflicts*

In this section, we explore the relationships between pastoral conflict in a cell and the weather and terrain-related variables previously described. For each ($100km \times 100km$) cell within a country, we investigate the association between the recorded values of these features and the occurrence of pastoral attacks. Our analysis seeks to answer the following question:

Given a weather/terrain variable *v*, is there a statistically significant difference in its mean value between cells where there is a pastoral conflict *(Class 1)* and no conflict *(Class 0)*?

*Weather Related Hypotheses Testing*

**Hypothesis 1** For each of the features maximum temperature at 2 meters (T2M_MAX), minimum temperature at 2 meters (T2M_MIN), relative humidity at 2 meters (RH2M), corrected total precipitation (PRECTOTCORR), maximum wind speed at 10 meters (WS10M_MAX), and minimum wind speed at 10 meters (WS10M_MIN): There is a statistically significant difference in the mean values of these variables between Classes 1 and 0.

Recall that each of the aforementioned meteorological factors was transformed into features using a histogram binning approach. This method involves segmenting the continuous range of each variable into 10 distinct bins, thereby quantifying the distribution of each variable's values within every 100 km cell grid. Each bin is represented as a separate feature, with the naming convention combining the feature name and the bin number. [3]

We formally compare the number of conflicts between the two classes using a linear model-based approach (two-sided Welch's *t*-test) between classes 0 and 1. [59] We report the mean difference along its 95% confidence interval (CI) and *P*-value between the number of attacks between the two classes. We then repeat the analysis for all collected features and in the respective countries. When the difference is statistically significant (i.e. $P < 0.05$), the 95% CI does not contain the null value of 0 (i.e. no difference). We address the challenge of multiple hypothesis testing using the Bonferroni correction, given the conditionally independent nature of the variables.

---

[3] For instance, as seen in Appendix A, the maximum temperature feature *T2M_MAX* is divided into 10 bins labeled *T2M_MAX1*, *T2M_MAX2*, ..., *T2M_MAX10*. This notation helps to systematically analyze each feature's impact on the likelihood of pastoralist conflicts occurring within these cells.

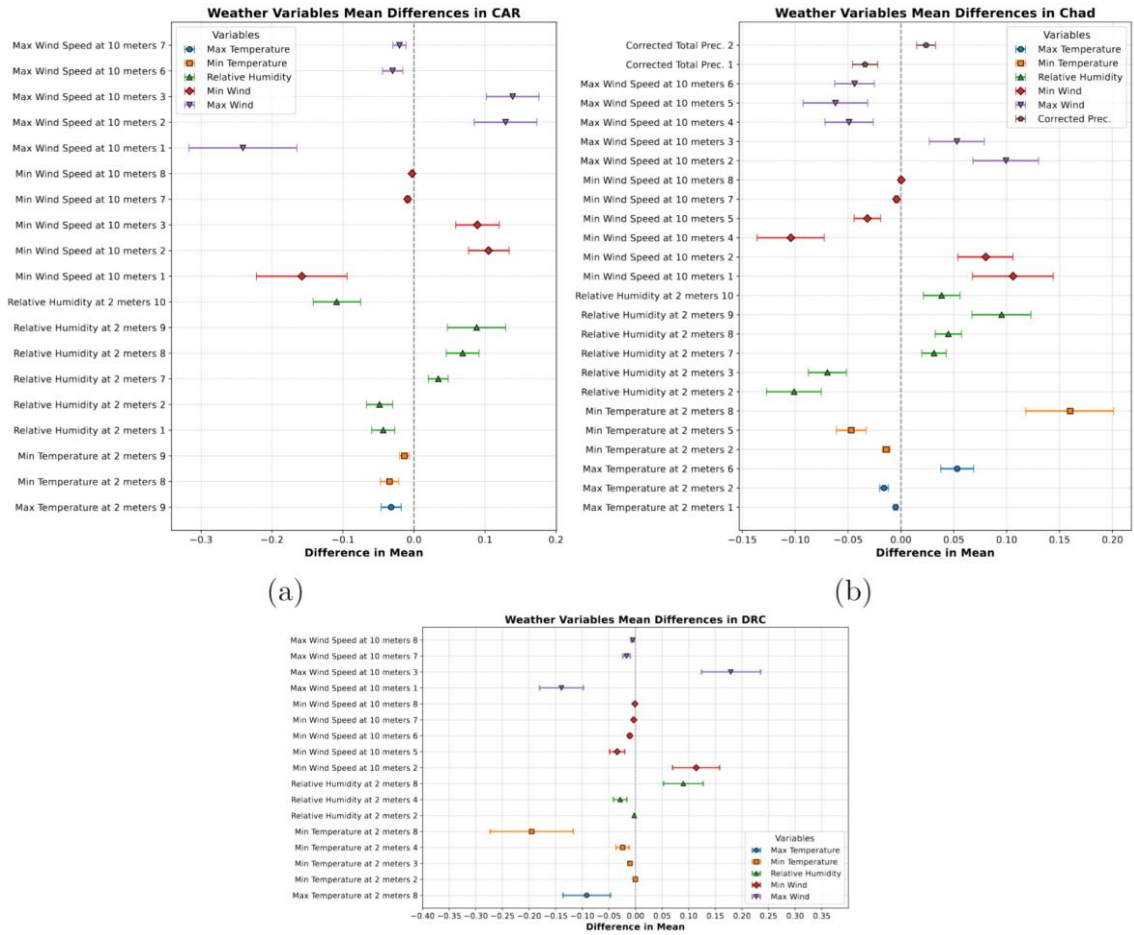

**Figure 4.** Forest Plot of Mean Differences with 95% Confidence Intervals for the Univariate Weather-Related Hypothesis for (a) CAR (b) Chad and (c) DRC.

***Results and Analysis of Hypothesis 1 for Cameroon:*** Table 20 in Appendix C presents the Bonferroni-corrected *P*-values for our analysis concerning Cameroon. Based on the data in the table, we conclude that there are no statistically significant differences in the mean values of the weather features between Class 1 (cells with pastoralist conflicts) and Class 0 (cells without conflicts).

***Results and Analysis of Hypothesis 1 for CAR:*** Table 21 in Appendix C displays the Bonferroni-corrected *P*-values from our analysis for CAR. Based on the results presented in the table as well as the forest plot showcased in Figure 4(a), we conclude that:

- There is a statistically significant higher occurrence of pastoral conflicts in Class 1 (cells with conflicts) for cells exhibiting the following meteorological features. These include:
    - Minimum temperature at 2 meters (bins 8, 9),
    - Maximum temperature at 2 meters (bin 9),
    - Relative humidity at 2 meters (bins 1, 2, 7, 8, 10),
    - Minimum wind speed at 10 meters (bins 1, 2, 3, 7, 8),
    - Maximum wind speed at 10 meters (bins 1, 2, 3, 6, 7).
- The statistical significance of these results is confirmed with Bonferroni-corrected *P*-values well below 0.05 in all noted cases.

Our results reveal that the mean differences between Class 1 and Class 0 for the weather-related variables of minimum and maximum temperature at 2 meters are negative (T2M_MIN/T2M_MAX). This indicates that these temperature values are lower for Class 1 and higher for Class 0, suggesting that regions with cooler air temperatures at 2 meters above the ground are more susceptible to pastoral conflicts. These findings support the hypothesis that areas with environmental conditions less conducive to vegetation growth are associated with an increased likelihood of pastoral conflicts in the CAR. Other weather factors, however, provide more surprising results that highlight the role that the geographical location within the country can have in the values that these environmental factors take and their role in the occurrence of pastoral conflicts. For example, our analysis of relative humidity at 2 meters (RH2M) provides further insights into the environmental factors influencing pastoral conflicts.

Specifically, the average value of RH2M in Class 1 (cells with pastoral conflict) when the mean difference is less than 0 is smaller than 0.0345, while in the scenario when the mean difference is greater than 0, the average of the RH2M values in Class 1 is greater than 0.11. In particular, the mean differences for RH2M1, RH2M2, and RH2M10 are negative, indicating that cells experiencing pastoral conflict (Class 1) have lower relative humidity compared to cells which did not experience pastoral conflict (Class 0), which could correspond to drier conditions and increased resource scarcity. Conversely, the mean differences for RH2M7, RH2M8, and RH2M9 are positive, indicating higher relative humidity in Class 1, which might correspond to more humid conditions potentially leading to waterlogging and reduced pasture quality. These results highlight the complex interplay between relative humidity and occurrence of pastoral conflict, suggesting that both extremely dry and excessively humid conditions can exacerbate tensions among pastoral communities.

In a similar manner, the maximum wind speed at 10 meters (WS10M) provides further insights. When the mean difference is less than 0, the average values for the maximum wind speed at 10 meters are less than 0.0475, indicating lower wind speeds during conflict periods. Conversely, when the mean difference is greater than 0, the average values for WS10M in Class 1 (cells with pastoral conflict) are significantly higher (greater than 0.2844). These results suggest that lower wind speeds may be associated with pastoral conflicts, potentially due to stagnant air leading to higher temperatures and lower humidity, which stress resources. Higher wind speeds, on the other hand, might be indicative of more dynamic weather conditions that could also contribute to pastoral conflicts in different ways, such as by affecting livestock or causing damage to resources.

***Results and Analysis of Hypothesis 1 for Chad:*** Table 22 in Appendix C displays the Bonferroni-corrected *P*-values from our analysis for Chad. Based on the results

presented as well as the forest plot in Figure 4(b), we can draw the following conclusions:

- There is a statistically significant higher occurrence of pastoral conflicts in Class 1 (cells with pastoral conflicts) for several meteorological features. These include:
    - Minimum temperature at 2 meters (bins 2, 5, 8),
    - Maximum temperature at 2 meters (bins 1, 2, 6),
    - Relative humidity at 2 meters (bins 2, 3, 7, 8, 9, 10),
    - Corrected total precipitation (bins 1 and 2),
    - Minimum wind speed at 10 meters (bins 1, 2, 4, 5, 7, 8),
    - Maximum wind speed at 10 meters (bins 2, 3, 4, 5, 6).
- The statistical significance of these results is confirmed with Bonferroni-corrected *P*-values well below 0.05 in all noted cases.

The analysis showed that the average value of the maximum temperature of air at 2 meters in Class 1 (cells with pastoral conflict) when the mean difference is greater than 0 is 0.218, while when it is less than 0, the average value is 0. Specifically, the mean differences for T2M_MAX1 and T2M_MAX2 are negative, indicating lower maximum temperatures at 2 meters in Class 1. This suggests that cooler air temperatures are linked to a higher number of pastoral conflicts, potentially because such temperatures are less conducive to vegetation growth. Conversely, the mean difference for T2M_MAX6 is positive, indicating higher maximum temperatures at 2 meters in Class 1. This could correspond to hotter conditions, which may also create stress on resources and contribute to pastoral conflicts.

Additionally, similar patterns were observed for other weather-related factors. For the minimum temperature at 2 meters, T2M_MIN2 and T2M_MIN5 have negative mean differences, with the average values being smaller than 0.0414, while T2M_MIN8 has a positive mean difference with an average value of 0.438. This further emphasizes the role of extreme temperature variations in pastoral conflict occurrence. Wind speed data also revealed that both lower and higher wind speeds at 10 meters can influence pastoral conflicts. Specifically, WS10M_MIN4, WS10M_MIN5, WS10M_MIN7, and WS10M_MIN8 show negative mean differences (with average values less than 0.018), while WS10M_MIN1 and WS10M_MIN2 show positive mean differences (with average values greater than 0.34).

For maximum wind speed at 10 meters, WS10M_MAX4, WS10M_MAX5, and WS10M_MAX6 have negative mean differences (with average values smaller than 0.22), while WS10M_MAX2 and WS10M_MAX3 have positive mean differences (with average values greater than 0.23). This indicates that varying wind speeds impact pastoral conflict occurrence in different ways. Precipitation data also highlighted significant differences, with PRECTOTCORR1 showing a negative mean difference, suggesting that cells with pastoral conflict (Class 1) are more likely to experience pastoral conflicts when the region receives less precipitation, perhaps because lower precipitation is linked to poorer soil and less vegetation growth. In contrast, PRECTOTCORR2 shows a positive mean difference.

***Results and Analysis of Hypothesis 1 for DRC:*** Table 23 in Appendix C displays the Bonferroni-corrected *P*-values from our analysis for DRC. Based on the results presented as well as the forest plot in Figure 4(c), we can draw the following conclusions:

- There is a statistically significant higher occurrence of pastoral conflicts in Class 1 (cells with conflicts) for several meteorological features. These include:
    - Maximum temperature at 2 meters (bin 8),
    - Relative humidity at 2 meters (bins 2, 4, 8),
    - Minimum wind speed at 10 meters (bins 2, 5, 6, 7, 8),
    - Maximum wind speed at 10 meters (bins 1, 3, 7, 8).
- The statistical significance of these results is confirmed with Bonferroni-corrected *P*-values well below 0.05 in all noted cases.

For the DRC, we showed that the mean differences between *Class 1* (pastoral conflict) and *Class 0* (no pastoral conflict) for the weather-related variables of minimum and maximum temperature at 2 meters (T2M_MIN/T2M_MAX) are negative. This indicates that these temperature values are lower in cells that experienced pastoral conflicts compared to cells that did not, suggesting that areas with cooler air temperatures at 2 meters above the ground are more prone to pastoral conflicts. These findings support the hypothesis that regions with environmental conditions that hinder vegetation growth are more likely to experience pastoral conflicts.

The rest of the data, similar to CAR and Chad, requires a more thorough interpretation due to the deeper connection and instability linked to geographical location within the country. Our results show that the mean differences for RH2M2 and RH2M4 are less than 0, with average values of less than 0.013 in Class 1, indicating that lower relative humidity is associated with cells that experience pastoral conflict. Conversely, RH2M8 has a positive mean difference, with an average value of 0.2814, suggesting that both extremely dry and extremely humid conditions may lead to pastoral conflicts.

For minimum wind speed at 10 meters, WS10M_MIN6, WS10M_MIN5, WS10M_MIN7, and WS10M_MIN8 have negative mean differences, with average values of 0.0018, 0.0122, 0.0, and 0.0, respectively, while WS10M_MIN2 has a positive mean difference with an average value of 0.3240. These findings suggest that both lower and higher minimum wind speeds are linked to cells that experience pastoral conflict.

Similarly, for maximum wind speed at 10 meters, WS10M_MAX1, WS10M_MAX7, and WS10M_MAX8 show negative mean differences, with average values of less than 0.0394, while WS10M_MAX3 has a positive mean difference with an average value of 0.3609. This indicates that both lower and higher maximum wind speeds are present in cells that experience pastoral conflict.

*Terrain Related Hypotheses Testing*

In this section, we discuss hypotheses related to terrain-related features.

***Hypothesis 2:*** For each of the following variables: leaf area indexed (LAI), greenness fraction (GRN), surface soil wetness (SSW), surface soil temperature (SST), and land evaporation (LNDEV), it is the case that there is a statistically significant difference in the mean values of these variables between Classes 1 and 0.

As in the case of our weather factors, the terrain-related factors were also transformed into features using a histogram-binning approach. We use the same statistical approach to address this hypothesis as in Hypothesis 1 and repeat the analysis for all the binning variations.

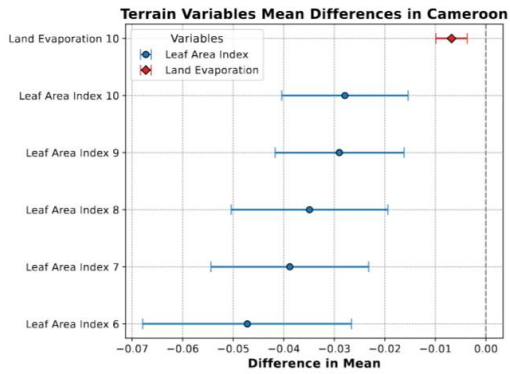
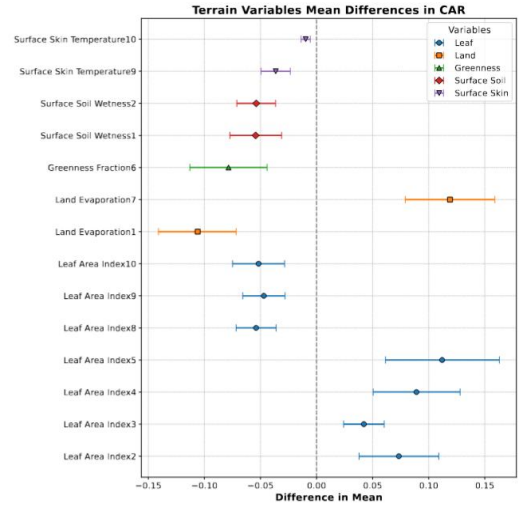
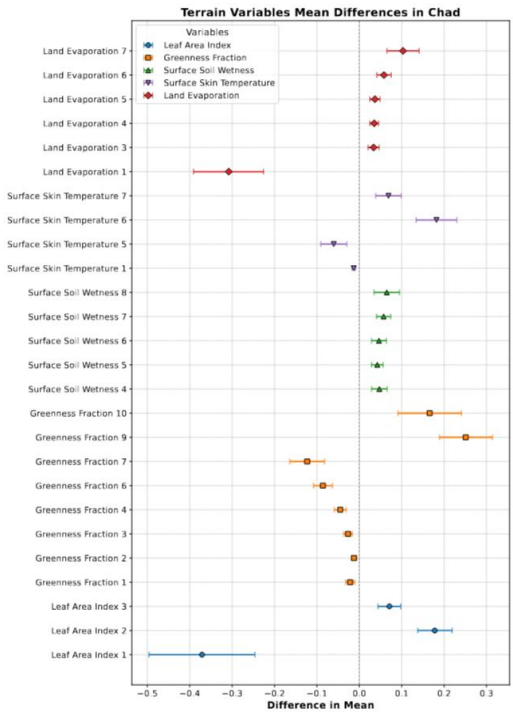
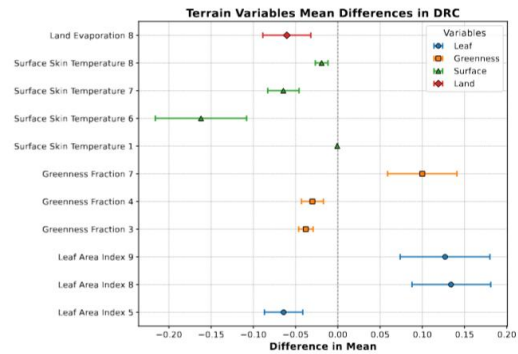

**Figure 5.** Forest Plot of Mean Differences with 95% Confidence Intervals for the Univariate Terrain Related Hypothesis for (a) Cameroon (b) CAR (c) Chad and (d) DRC.

***Results and Analysis of Hypothesis 2 for Cameroon***: Table 20 in Appendix C presents the Bonferroni-corrected *P*-values for our analysis concerning Cameroon. Based on the data in the table as well as the forest plot showcased in Figure 5(a), we conclude that:

- There is a statistically significant higher occurrence of pastoral conflicts in Class 1 (cells with conflicts) for several terrain-related features. These include:
    - Leaf area indexed (bins 6, 7, 8, 9, 10),
    - Land evaporation (bin 10).
- There is a negative association between pastoral conflicts and the terrain related variables of leaf area indexed and land evaporation.
- The statistical significance of these results is confirmed with Bonferroni-corrected P- values well below 0.05 in all noted cases.

The result of testing of Hypothesis 2 for Cameroon requires a nuanced interpretation. Our findings indicate that the mean differences between Class 1 (cells with pastoral conflict) and Class 0 (cells without pastoral conflict) for the terrain-related variables are negative. Specifically, the recorded values of the leaf area indexed and land evaporation indicators, in the respective range of values, are lower for Class 1 and higher for Class 0. This suggests that areas with lower vegetation and land evaporation are more prone to experiencing pastoral conflicts. These results support the hypothesis that the depletion of vegetation and a lower percentage of land evaporation are associated with a higher likelihood of pastoral conflicts in Cameroon.

***Results and Analysis of Hypothesis 2 for CAR:*** Table 21 in Appendix C displays the Bonferroni-corrected *P*-values for the CAR analysis. Based on the results presented in the table as well as the forest plot showcased in Figure 5(b), we can draw the following conclusions:

- There is a statistically significant higher occurrence of pastoral conflicts in Class 1 (cells with conflicts) for several terrain-related features. These include:

- Leaf area indexed (bins 2, 3, 4, 5, 8, 9, 10)
    - Greenness fraction (bin 6),
    - Surface soil wetness (bins 1, 2),
    - Land evaporation (bins 1, 7),
    - Surface skin temperature (bins 9, 10).
- The statistical significance of these results is confirmed with Bonferroni-corrected *P*-values well below 0.05 in all noted cases.

The results reveal that the mean differences between Class 1 and Class 0 across terrain-related variables such as greenness fraction, surface soil wetness, and surface skin temperature are negative. Specifically, values for these variables are lower in Class 1 (cells with pastoral conflict) and higher in Class 0 (cells without). This suggests that regions with diminished vegetation and soil less conducive to supporting plant life are more susceptible to pastoral conflicts. These findings support the hypothesis that areas with scarcer resources or less fertile soil are more likely to experience pastoral conflicts.

Additionally, our analysis of leaf area index (LAI) shows that the mean differences for LAI8, LAI9, and LAI10 are negative, with average values of 0.0 in Class 1, indicating lower vegetation density in conflict regions. Conversely, LAI2, LAI3, LAI4, and LAI5 have positive mean differences, with average values of 0.2182, 0.1616, 0.1975, and 0.1967, respectively, suggesting that higher vegetation density is also associated with pastoral conflicts, possibly due to other underlying factors such as competition for resources.

For land evaporation (LNDEV), the mean difference for LNDEV1 is negative, with average values of 0.0204 and while the mean difference for LNDEV2 is positive with an average value of 0.0314 in Class 1, respectively. This indicates that areas with lower land evaporation are associated with higher conflict occurrence, further emphasizing the link between the area's evaporation rate, vegetation density and pastoral conflicts.

In conclusion, the terrain-related variables, including leaf area index, greenness fraction, surface soil wetness, surface skin temperature, and land evaporation, highlight the intricate relationship between environmental conditions and pastoral conflicts. These findings underscore the importance of addressing environmental degradation and resource scarcity to mitigate conflict risks in the CAR.

***Results and Analysis of Hypothesis 2 for Chad***: Table 22 in Appendix C displays the Bonferroni-corrected *P*-values from our analysis for Chad. Based on the results presented as well as the forest plot showcased in Figure 5(c), we can draw the following conclusions:

- There is a statistically significant higher occurrence of pastoral conflicts in Class 1 (cells with conflicts) for several terrain-related features. These include:
    - Leaf area indexed (bins 1, 2, 3)
    - Greenness fraction (bins 1, 2, 3, 4, 6, 7, 9, 10),
    - Surface skin temperature (bins 1, 5, 6, 7),
    - Land evaporation (bin 1, 3, 4, 5, 6, 7).
- The statistical significance of these results is confirmed with Bonferroni-corrected *P*-values well below 0.05 in all noted cases.

The results show that the mean difference between Class 1 (cells with pastoral conflict) and Class 0 (cells without) for terrain-related variables like surface soil wetness are positive. Specifically, values for these variables are higher in Class 1 than in Class 0. This indicates that cells with soil that is more saturated and, therefore, more conducive to supporting plant health and growth, are less likely to experience pastoral conflicts.

Additionally, for the leaf area index (LAI), the mean differences for LAI2 and LAI3 are positive, with average values of 0.2601 and 0.1195, respectively, in Class 1. Conversely, LAI1 has a negative mean difference with an average value of 0.3664, suggesting that cells with denser vegetation are more prone to pastoral conflicts, possibly due to competition for resources.

For surface skin temperature (SST), SST6 and SST7 have positive mean differences with average values of 0.3712 and 0.2234, respectively, indicating that cells with higher temperatures are more likely to experience pastoral conflicts. On the other hand, SST1 and SST5 have negative mean differences with average values of 0.0 and 0.0554, respectively, suggesting that cooler cells may also be susceptible to pastoral conflicts under certain conditions.

In terms of land evaporation (LNDEV), LNDEV3, LNDEV4, LNDEV5, LNDEV6, and LNDEV7 show positive mean differences, with average values of 0.0721, 0.0673, 0.0697, 0.0936, and 0.1748, respectively. This indicates that cells with higher rate of evaporation are associated with a higher likelihood of pastoral conflict. Conversely, LNDEV1 has a negative mean difference with an average value of 0.3724, suggesting that lower land evaporation in cells might lead to experiencing more conflicts.

Regarding greenness fraction (GRN), GRN9 and GRN10 have positive mean differences, with average values of 0.4062 and 0.308, respectively, indicating that cells in Chad with higher greenness are more likely to have pastoral conflicts. However, GRN1, GRN2, GRN3, GRN4, GRN6, and GRN7 have negative mean differences, with

GRN6 and GRN7 showing average values of 0.0064 and 0.0669, respectively, suggesting that less green areas in Chad are also prone to conflicts.

***Results and Analysis of Hypothesis 2 for DRC:*** Table 23 in Appendix C displays the Bonferroni-corrected *P*-values from our analysis for DRC. Based on the results presented in the table as well as the forest plot showcased in Figure 5(d), we can draw the following conclusions:

- There is a statistically significant higher occurrence of pastoral conflicts in Class 1 (cells with conflicts) for several terrain-related features. These include:
    - Leaf area indexed (bins 5, 8, 9)
    - Greenness fraction (bins 3, 4, 7),
    - Land evaporation (bin 8).
    - Surface skin temperature (bins 1, 6, 7, 8).
- The statistical significance of these results is confirmed with Bonferroni-corrected *P*-values well below 0.05 in all noted cases.

The findings indicate that the mean differences between Class 1 (cells with pastoral conflict) and Class 0 (cells without) for terrain-related variables like surface skin temperature and land evaporation are negative. Specifically, the values for these variables are lower in Class 1 and higher in Class 0. This suggests that cells with reduced thermal energy emitted from the earth's surface and consequently lower rates of land evaporation are more vulnerable to pastoral conflicts.

For the leaf area index (LAI), the mean difference for LAI5 is negative, with an average value of 0.0613 in Class 1. Conversely, LAI8 and LAI9 have positive mean differences, with average values of 0.3222 and 0.2036, respectively, indicating that cells with higher vegetation density might be associated with pastoral conflicts, potentially due to competition for resources or other underlying factors.

Regarding greenness fraction (GRN), the mean differences for GRN3 and GRN4 are negative, with average values of 0.0186 and 0.0323, respectively, in Class 1. On the other hand, GRN7 has a positive mean difference with an average value of 0.0323, suggesting that both lower and higher greenness can be linked to pastoral conflict.

These findings underscore the complexity of the relationship between environmental conditions and pastoral conflicts in the DRC. It highlights the importance of considering a range of terrain-related variables, including vegetation density, surface temperature, and land evaporation, when addressing conflict risks in the region.

*Machine Learning Based Analysis*

In this section, we present the findings from our machine learning analysis aimed at predicting the occurrence of pastoral conflicts in each cell in Cameroon, CAR, Chad, and the DRC. It is important to clarify that the objective of this paper is not to introduce new machine learning methodologies, but rather to apply existing techniques to better understand the risk of pastoral conflict in cells in these countries. Additionally, to examine the impact of spatial resolution on our predictions, we constructed similar datasets with grid granularities set to $(75km \times 75km)$ and $(50km \times 50km)$, in addition to the initially mentioned granularity of $(100km \times 100km)$ granularity used in the statistical analysis described previously in the paper.

*Predictive Performance:* To carry out our country-specific experiments, our data for each country consisted of triples $(s, \vec{f_s}, \vec{dv_s})$ where $s$ is a cell in the country's grid, $\vec{f_s}$ is the feature vector of length 120 associated with $s$, and $\vec{dv_s}$ is the binary dependent variable. The feature vector $\vec{f_s}$ includes the 120 features detailed in S1 Appendix A,

Table 7. The value of the dependent variable $\overrightarrow{dv_s}$ is set to 1 if a pastoral conflict occurred in cell $s$ during the study timeframe --- otherwise, it is set to 0. We applied eight well-known machine learning classifiers to this data: Random Forest, Decision Trees, AdaBoost, Logistic Regression, Linear SVM, Gaussian Naive Bayes, Multi-Layer Perceptron, and Deep Neural Networks. The last two are deep learning classifiers. The Precision, Recall, F1-Score, and Area under the Receiver Operating Characteristic Curve (AUC) for the best-performing classifiers across the four countries and their respective granularities are shown in Tables 2, 3, and 4. Both AUC and F1-Scores serve as single performance metrics, with the latter combining Precision and Recall.

We detail the outcomes of the country-specific experimental assessments of predictive performance at all three granularities:

*100km Grid Cell Granularity:*

| Country | Best Classifier | Precision | Recall | F1-Score | AUC |
| --- | --- | --- | --- | --- | --- |
| Cameroon | Decision Tree conflict class | 0.83 | 1 | 0.91 | 0.99 |
| CAR | Linear SVM conflict class | 1 | 0.57 | 0.73 | 0.82 |
| Chad | Linear SVM conflict class | 0.4 | 0.5 | 0.44 | 0.83 |
| DRC | AdaBoost conflict class | 0.71 | 0.83 | 0.77 | 0.98 |

**Table 2.** Best performing classifier metrics per country for predicting pastoral conflicts at 100 km cell granularity.

*75km Grid Cell Granularity:*

| Country | Best Classifier | Precision | Recall | F1-Score | AUC |
|---|---|---|---|---|---|
| Cameroon | MLP conflict class | 0.5 | 0.5 | 0.5 | 0.83 |
| CAR | Logistic Regression conflict class | 0.75 | 0.69 | 0.72 | 0.95 |
| Chad | Linear SVM conflict class | 0.71 | 0.56 | 0.63 | 0.88 |
| DRC | MLP conflict class | 0.6 | 0.55 | 0.57 | 0.9 |

**Table 3.** Best performing classifier metrics per country for predicting pastoral conflicts at 75 km cell granularity.

*50km Grid Cell Granularity:*

| Country | Best Classifier | Precision | Recall | F1-Score | AUC |
|---|---|---|---|---|---|
| Cameroon | MLP conflict class | 0.57 | 0.8 | 0.67 | 0.89 |
| CAR | Logistic Regression conflict class | 0.55 | 0.32 | 0.4 | 0.73 |
| Chad | Linear SVM conflict class | 0.44 | 0.31 | 0.36 | 0.64 |
| DRC | MLP conflict class | 0.53 | 0.56 | 0.54 | 0.89 |

**Table 4.** Best performing classifier metrics per country for predicting pastoral conflicts at 50 km cell granularity.

The performance variability of different classifiers across countries, as detailed in Tables 2, 3, and 4, is affected by class imbalance. Notably, as grid granularity becomes finer—transitioning from $(100km \times 100km)$ to $(50km \times 50km)$, the number of data points per cell decreases. The choice of the best classifier changes with granularity, suggesting that no single predictive model universally outperforms others across different spatial resolutions and contexts. This variability indicates that local data characteristics and pastoral conflict dynamics differ significantly between countries, influencing the models' effectiveness.

As grid granularity becomes finer, the metrics of the best-performing classifiers shift. For example, in the Central African Republic, the highest F1-Score at a ($100km \times 100km$) granularity was achieved using a Linear SVM (0.73). However, at a ($50km \times 50km$) granularity, the best performance was obtained with Decision Trees, resulting in a significantly lower F1-Score (0.40). This shift demonstrates that finer granularity does not uniformly enhance performance and may introduce greater noise or reduce differentiation between cells, thereby complicating predictive accuracy.

In particular, for Cameroon, we observe that at the ($100km \times 100km$) granularity, the Decision Tree classifier performed exceptionally well with a Precision of 0.83, Recall of 1.00, and F1-Score of 0.91, indicating strong predictive power at this resolution. At a ($75km \times 75km$) granularity, the performance dropped with an MLP classifier yielding balanced but lower scores. At the ($50km \times 50km$) granularity, the Neural Network classifier showed a slightly increased performance compared to the ($75km \times 75km$) granularity with an F1-Score of 0.67, reflecting the challenges of finer granularity.

For the Central African Republic, at the ($100km \times 100km$) granularity, the Linear SVM classifier performed well with an F1-Score of 0.73. This performance remained consistent at the ($75km \times 75km$) granularity with the Logistic Regression classifier but declined at the ($50km \times 50km$) granularity, with the Decision Tree classifier scoring an F1-Score of 0.40, indicating that finer granularity may introduce noise and reduce predictive accuracy. In Chad, the best performance was observed at the ($75km \times 75km$) granularity with an F1-Score of 0.63. This performance decreased again at the ($50km \times 50km$) granularity, where the Deep Neural Network recorded a lower F1-Score of 0.36. Lastly, in the Democratic Republic of Congo, the best-performing classifier at a ($100km \times 100km$) granularity was AdaBoost with an F1-

Score of 0.77. This score gradually decreased as the granularity of the cells became finer.

Appendix B provides a comprehensive breakdown of the predictive performance of all tested classifiers across various countries and granularity.

*Multivariate Hypotheses Using Decision Trees*

In addition to predictions generated by our suite of classifiers, we used Decision Tree Classifiers because of their easy interpretability, which subsequently inspired us to formulate multivariate hypotheses for deeper understanding of the data.

Figure 6 displays the decision tree we learned for Cameroon at a grid granularity of $(100km \times 100km)$. The bold blue line in Figure 6 illustrates the following hypothesis:

Let $S$ denote the set of cells within a country satisfying these conditions: the total number of pastoral conflicts within a distance of one cell from the cell in question exceeds 9.5, and the probability that the cell's surface soil wetness falls within the [30,40] interval is below 0.09. Among the 11 cells meeting these criteria, 10 experienced pastoral conflicts. Based on this, we hypothesized that cells in $S$ have a significantly higher likelihood of experiencing a pastoral conflict compared to cells in $\bar{S}$. This method of hypothesis generation is consistent across all the decision trees we have constructed for each country at a cell granularity of $(100km \times 100km)$.

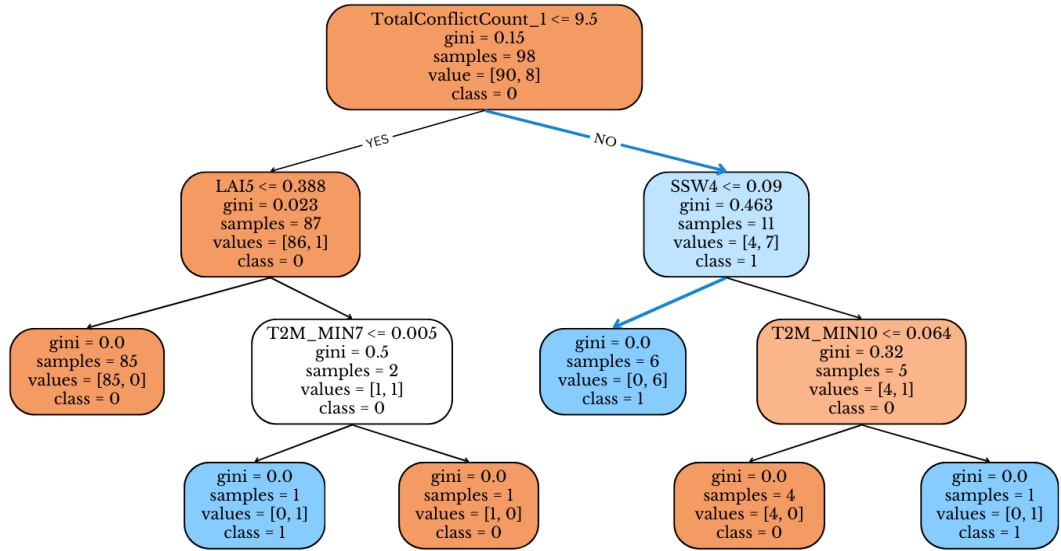

**Figure 6.** Learned Decision Tree for Cameroon (100km×100km) grid cell granularity with highlighted path for hypothesis 3.

### *Statistical Inference of Hypotheses in Pastoral Conflicts Inspired by Machine Learning Results*

We formally address these types of hypotheses using a maximum likelihood-based statistical framework. The paths we select in our decision trees are highlighted in the hypotheses below. We first investigate the hypotheses derived from the example shown in Figure 6.

*Cameroon*

***Hypothesis 3:*** Let $S$ be the set of 100 km cells in Cameroon that satisfy the following logical "and" condition. Each cell $s \: \epsilon \: S$ must be such that:

(1) There are more than 9.5 pastoral conflicts occurring within a neighboring distance of one cell from the cell in question.

(2) The probability that the cell's surface soil wetness is within the [30, 40] interval is less than 0.09.

Let $\bar{S}$ be the set of all other cells in Cameroon (i.e. those that do not satisfy the above condition). We hypothesize that cells in $S$ are more likely to experience a pastoral conflict than cells in $\bar{S}$.

There were 12 cells in $S$ and 11 of them experienced pastoral conflicts. Table 5 shows the contingency table for testing the hypothesis. We use an odds ratio (OR) as the summary measure to represent how many times pastoral conflicts occurred in $S$ relative to $\bar{S}$. Recall that

- $OR > 1$ indicates pastoral conflicts in set $S$ is $OR$ fold higher than $\bar{S}$.
- $OR < 1$ indicates pastoral conflicts in set $S$ is $OR$ fold lower than than $\bar{S}$ (or equivalently, frequency $S$ is $\frac{1}{OR}$ times of $\bar{S}$).
- $OR = 1$, the null value, indicates there is no fold difference between set $S$ and $\bar{S}$.

We used Fisher's exact test [60] to formally evaluate this hypothesis and quantify the uncertainties. When the fold difference between sets $S$ and $\bar{S}$ is statistically significant, the 95% $CI$ of the odds ratio (OR) will contain the null value of 1 and the associated $P$-value would be less than the 0.05 threshold.

| Category | Set S | Set $\bar{S}$ |
|---|---|---|
| Attack = 1 | 11 (91.67%) | 2 (1.56%) |
| Attack = 0 | 1 (8.33%) | 126 (98.44%) |

**Table 5.** Example of contingency table setup for testing Hypothesis 3.

We similarly set up a contingency table for each of the remaining hypotheses from decision trees of countries Chad, CAR and the DRC. Table 6 displays the results of all Fisher's exact tests across the four countries. We find that in general:

- All OR point estimates were positive in direction and distant to the null value of 1.
- The 95% confidence interval (CI) estimates did not contain the null value of 1.
- All Fisher's exact test *P*-values were well below $10^{-3}$.

These results strongly support Hypothesis 3 for Cameroon.

| Country | Hypothesis | Attacks Set S | | Attacks Set $\bar{S}$ | | Odds Ratio | 95% CI, lower | 95% CI, upper | P-value |
|---|---|---|---|---|---|---|---|---|---|
| | | Count | % | Count | % | | | | |
| Cameroon | Hyp.3 | 11 | 91.67 | 2 | 1.56 | 693 | 58.13 | 8261.12 | 1.36E-13 |
| CAR | Hyp.4 | 14 | 87.5 | 24 | 15 | 39.67 | 8.471 | 185.74 | 4.66E-09 |
| CAR | Hyp.5 | 8 | 72.73 | 30 | 18.18 | 12 | 3.005 | 47.92 | 2.47E-04 |
| Chad | Hyp.6 | 5 | 62.5 | 22 | 10.33 | 14.47 | 3.236 | 64.71 | 8.25E-04 |
| Chad | Hyp.7 | 10 | 71.43 | 17 | 8.21 | 27.94 | 7.916 | 98.63 | 9.98E-08 |
| Chad | Hyp.8 | 5 | 71.43 | 22 | 10.28 | 21.82 | 3.993 | 119.21 | 3.38E-04 |
| DRC | Hyp.9 | 5 | 83.33 | 25 | 5 | 95 | 10.69 | 844.08 | 3.02E-06 |
| DRC | Hyp.10 | 11 | 78.57 | 19 | 3.86 | 91.28 | 23.51 | 354.39 | 1.43E-12 |

**Table 6.** Results of Statistical Evaluation of the Decision Tree Hypotheses.

*CAR*

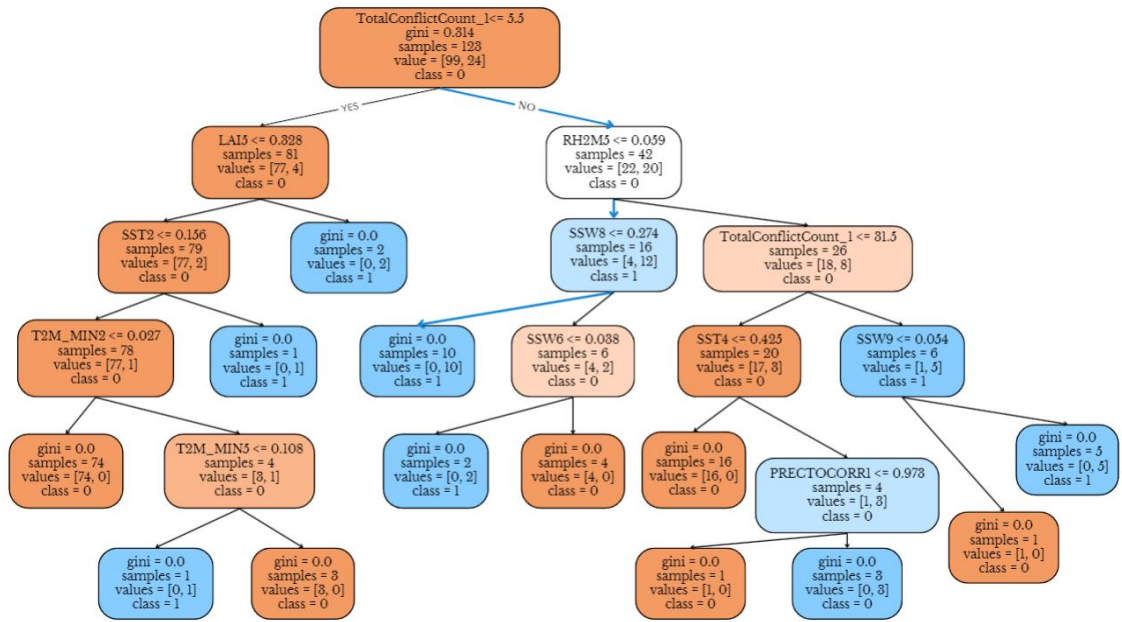

**Figure 7.** Learned Decision Tree for CAR, 100 km grid cell granularity with highlighted path for hypothesis 4.

***Hypothesis 4:*** (as visualized in blue in Fig. 7) Let *S* represent the set of 100 km cells in CAR that meet the following criteria:

(1) the total number of pastoral conflicts within a neighboring distance of one cell from the cell under study exceeds 5.5,

(2) the probability that the cell's relative humidity at two meters is within the [40, 50] interval is below 0.059 and

(3) the probability that the cell's surface soil wetness is within the [70, 80] interval is below 0.274.

Our hypothesis is that cells in *S* are more likely to experience a pastoral conflict than cells in $\bar{S}$. As shown in Table 6, the frequency of conflicts in *S* is over 21 times (95%CI = 8.47 − 185.74, $P = 4.66E - 9$), suggesting that this hypothesis is valid.

***Hypothesis 5:*** (as visualized in Appendix D, Fig. 10) Let S be the set of ($100km \times 100km$) cells in CAR that meet the following criteria:

(1) the total number of pastoral conflicts within a neighboring distance of one cell from the cell under study exceeds 5.5,

(2) the probability that the cell's relative humidity at two meters falls within the [40, 50] interval exceeds 0.059 and

(3) the total number of pastoral conflicts within a neighboring distance of one cell from the cell under study exceeds 31.5.

Our hypothesis is that cells in $S$ are more likely to experience a pastoral conflict than cells in $\bar{S}$.

As shown in Table 6, the frequency of pastoral conflicts in $S$ is over 15 times (95%CI = $3.01 - 47.92$, $P = 2.47E - 3$), suggesting that this hypothesis is valid.

*Chad*

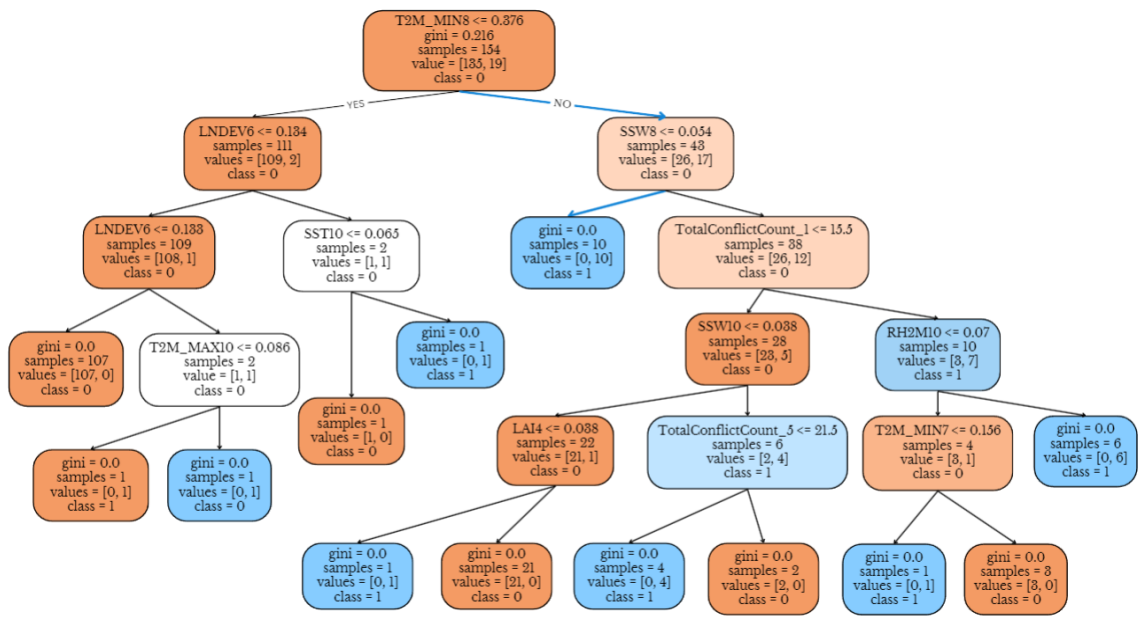

**Figure 8.** Learned Decision Tree for Chad, 100 km grid cell granularity with highlighted path for hypothesis 6.

***Hypothesis 6:*** (as visualized in blue in Fig. 8) Let $S$ represent the set of $(100km \times 100km)$ cells in Chad meeting the following criteria:

(1) the probability that the cell's minimum temperature at two meters falls within the [70, 80] range exceeds 0.376 and

(2) the probability that the cell's surface soil wetness is within the [70, 80] interval is below 0.054.

We hypothesize that cells in $S$ are significantly more likely to experience a pastoral conflict compared to cells in $\bar{S}$.

We formally determined that the frequency of pastoral conflicts in set $S$ is over 20 times (95%CI = 3.24 − 64.71) of set $\bar{S}$ with a highly significant $P$-value of $8.25E − 4$.

***Hypothesis 7:*** (as visualized in Appendix D, Fig. 11) Let $S$ represent the set of $(100km \times 100km)$ cells in Chad that satisfy the following conditions:

(1) the probability that the cell's minimum temperature at two meters falls within the [70, 80] interval exceeds 0.376 and

(2) the probability that the cell's surface soil wetness is within the [70, 80] interval exceeds 0.054 and

(3) the total number of pastoral conflicts occurring within a neighboring distance of one cell from the cell under study exceeds 15.5.

Our hypothesis is that cells in $S$ are significantly more likely to experience pastoral conflicts compared to cells in $\bar{S}$.

As shown in Table 6, the frequency of pastoral conflicts in $S$ is over 12 times (95%CI = 7.92 − 98.63, $P = 9.98E − 8$), suggesting that this hypothesis is valid.

***Hypothesis 8:*** (as visualized in Appendix D, Fig. 12) Let $S$ denote the set of ($100km \times 100km$) cells in Chad that meet the following criteria:

(1) the probability that the cell's minimum temperature at two meters is within the [70, 80] interval exceeds 0.376,

(2) the probability that the cell's surface soil wetness is within the [70, 80] interval exceeds 0.054,

(3) the total number of pastoral conflicts within a neighboring distance of one cell from the cell under study is less than 15.5 and

(4) the probability that the cell's surface soil wetness falls within the [90, 100] interval exceeds 0.038.

Our hypothesis is that cells in $S$ are more likely to experience a pastoral conflict than cells in $\bar{S}$.

As shown in Table 6, the frequency of pastoral conflicts in $S$ is over 29 times (95%CI = 3.99 − 119.21, $P = 3.38E − 4$), strongly support our hypothesis.

*DRC*

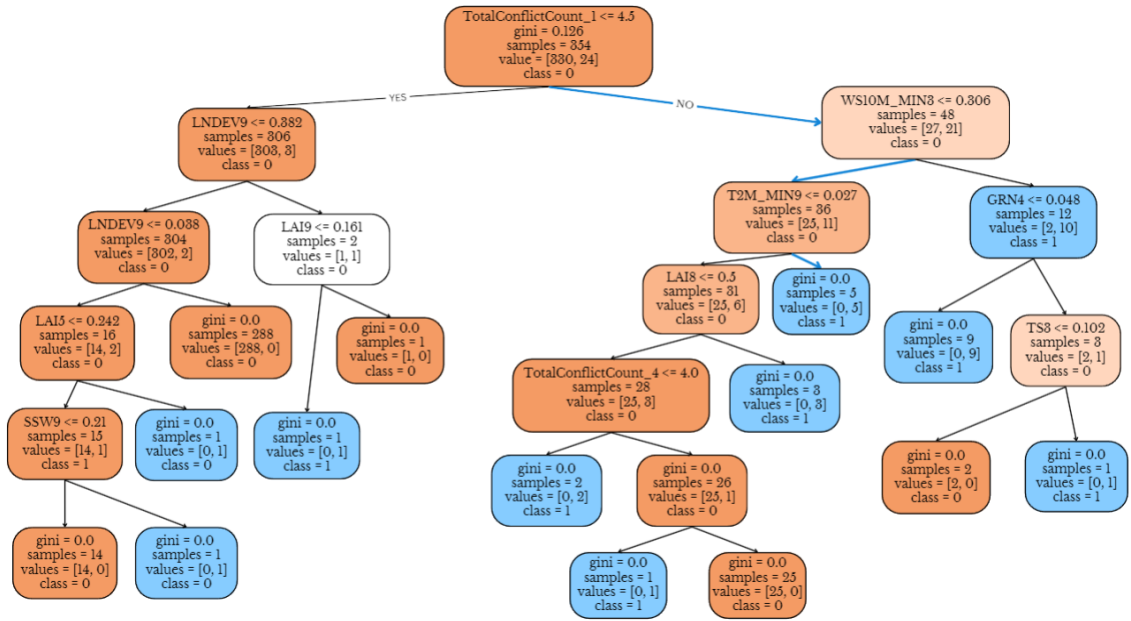

**Figure 9.** Learned Decision Tree for the DRC, 100 km grid cell granularity with highlighted path for hypothesis 9.

***Hypothesis 9:*** (as visualized in blue in Fig. 9) Let $S$ be the set of $(100km \times 100km)$ cells in the DRC that satisfy the following criteria:

(1) the total number of pastoral conflicts within a neighboring distance of one cell from the cell under study exceeds 4.5,

(2) the probability that the cell's minimum wind speed at ten meters falls within the [20, 30] interval is below 0.306 and

(3) the probability that the cell's minimum temperature at two meters falls within the [80, 90] interval exceeds 0.027.

Our hypothesis is that cells in $S$ are far more likely to experience pastoral conflicts than cells in $\bar{S}$.

As shown in Table 6, the frequency of pastoral conflicts in S is over 78 times (95%CI = 10.69 − 844.08, $P = 3.02E − 6$), supporting this hypothesis.

***Hypothesis 10:*** (as visualized in Appendix D, Fig. 13) Let $S$ be the set of $(100km \times 100km)$ cells in the DRC that satisfy the following logical condition:

(1) the total number of pastoral conflicts within a neighboring distance of one cell from the cell under study exceeds 4.5 and

(2) the probability that the cell's minimum wind speed at ten meters falls within the [20, 30] interval exceeds 0.306.

Our hypothesis is that cells in $S$ are more likely to experience a pastoral conflict than cells in $\bar{S}$. As shown in Table 6, the frequency of pastoral conflicts in $S$ is over 15 times (95%CI = 23.51 − 354.39, $P = 1.43E − 12$), supporting this hypothesis.

The statistical evidence presented in Table 6 supports the conclusion that cells within the country satisfying the conditions to be in set $S$ are far more likely to experience pastoral conflicts than cells that do not.

**Conclusion**

Our research demonstrates how statistically linked environmental risk factors of conflict vary across countries, suggesting "one size fits all" models are insufficient for understanding these dynamics. Even within a country, factors can be highly nuanced and delicate. We conclude that overgeneralized approaches to mapping pastoral conflict risk across a large number of countries may not lead to good policy recommendations for what a specific country should do.

Our statistical analysis provides a country-specific pastoral conflict risk estimate: one that assigns a probability of pastoral conflict to each $(100km \times 100km)$ cell in that country. Decision-makers can refer to this study to gather insight on which areas have the highest risk of pastoral conflict at specific times and use that knowledge to take steps to avert - or at least reduce the intensity of - such conflicts. In cells where pastoral conflict is linked to dry conditions during certain months, setting up appropriate canal systems and irrigation systems that harness excess water in other locations might help. In cells where pastoral conflict is linked to wet conditions during certain months, actors can consider constructing strategically located reservoirs that distribute water evenly from centralized zones to avoid competition over direct access methods might be used to set up reservoirs that are strategically located in places so that pastoralists seeking to access those reservoirs do not cross paths where other farmers and herders might compete for the same resource. Questions for future research include identifying "optimal" locations where to set up such facilities (both for livestock and for farming).

We recognize that these kinds of interventions can take years to develop, even if the political will to do so is present. In the meantime, lower sophistication interventions may still be useful. For example, these risk maps can support coordination strategies by directing pastoralists to routes and locations where the risk of conflicts is lower Such interventions could transmit this information via mobile phones capable of receiving SMSs that are compatible with the technology and mobile networks pastoralists are connected to in this region. Systems such as Maano in Uganda, iCow in Kenya, KokoLink in Ghana, Modisar in Botswana, all provide information to farmers already and have achieved decent penetration in the market. A system like Google Maps for herders, but with audio-visual user interfaces and the ability to host information in local dialects could help direct pastoralists to relatively low-risk areas. Whether an app or otherwise, this can spark inspiration for meaningful next steps on delivering information to stakeholders directly via creative public-private partnerships between telecom firms in Africa, big tech firms such as Google and Meta, local, regional, and national governments, and organizations such as the UN.

**Appendix A:** List of Features Associated with Each Grid Cell

**Table 7.** Independent features calculated for every cell record and their description.

| Independent Feature | Feature Definition |
|---|---|
| LAI1 | The probability that the cell's leaf area index is within the [0, 10] interval. |
| LAI2 | The probability that the cell's leaf area index is within the [10, 20] interval. |
| LAI3 | The probability that the cell's leaf area index is within the [20, 30] interval. |
| LAI4 | The probability that the cell's leaf area index is within the [30, 40] interval. |
| LAI5 | The probability that the cell's leaf area index is within the [40, 50] interval. |
| LAI6 | The probability that the cell's leaf area index is within the [50, 60] interval. |
| LAI7 | The probability that the cell's leaf area index is within the [60, 70] interval. |
| LAI8 | The probability that the cell's leaf area index is within the [70, 80] interval. |
| LAI9 | The probability that the cell's leaf area index is within the [80, 90] interval. |
| LAI10 | The probability that the cell's leaf area index is within the [90, 100] interval. |
| GRN1 | The probability that the cell's greenness fraction is within the [0, 10] interval. |
| GRN2 | The probability that the cell's greenness fraction is within the [10, 20] interval. |
| GRN3 | The probability that the cell's greenness fraction is within the [20, 30] interval. |
| GRN4 | The probability that the cell's greenness fraction is within the [30, 40] interval. |
| GRN5 | The probability that the cell's greenness fraction is within the [40, 50] interval. |
| GRN6 | The probability that the cell's greenness fraction is within the [50, 60] interval. |
| GRN7 | The probability that the cell's greenness fraction is within the [60, 70] interval. |
| GRN8 | The probability that the cell's greenness fraction is within the [70, 80] interval. |
| GRN9 | The probability that the cell's greenness fraction is within the [80, 90] interval. |
| GRN10 | The probability that the cell's greenness fraction is within the [90, 100] interval. |
| SSW1 | The probability that the cell's surface soil wetness is within the [0, 10] interval. |
| SSW2 | The probability that the cell's surface soil wetness is within the [10, 20] interval. |
| SSW3 | The probability that the cell's surface soil wetness is within the [20, 30] interval. |
| SSW4 | The probability that the cell's surface soil wetness is within the [30, 40] interval. |
| SSW5 | The probability that the cell's surface soil wetness is within the [40, 50] interval. |
| SSW6 | The probability that the cell's surface soil wetness is within the [50, 60] interval. |
| SSW7 | The probability that the cell's surface soil wetness is within the [60, 70] interval. |
| SSW8 | The probability that the cell's surface soil wetness is within the [70, 80] interval. |
| SSW9 | The probability that the cell's surface soil wetness is within the [80, 90] interval. |
| SSW10 | The probability that the cell's surface soil wetness is within the [90, 100] interval. |
| SST1 | The probability that the cell's surface skin temperature is within the [0, 10] interval. |
| SST2 | The probability that the cell's surface skin temperature is within the [10, 20] interval. |
| SST3 | The probability that the cell's surface skin temperature is within the [20, 30] interval. |
| SST4 | The probability that the cell's surface skin temperature is within the [30, 40] interval. |
| SST5 | The probability that the cell's surface skin temperature is within the [40, 50] interval. |
| SST6 | The probability that the cell's surface skin temperature is within the [50, 60] interval. |
| SST7 | The probability that the cell's surface skin temperature is within the [60, 70] interval. |
| SST8 | The probability that the cell's surface skin temperature is within the [70, 80] interval. |
| SST9 | The probability that the cell's surface skin temperature is within the [80, 90] interval. |
| SST10 | The probability that the cell's surface skin temperature is within the [90, 100] interval. |
| LNDEV1 | The probability that the cell's land evaporation rate is within the [0, 10] interval. |
| LNDEV2 | The probability that the cell's land evaporation rate is within the [10, 20] interval. |
| LNDEV3 | The probability that the cell's land evaporation rate is within the [20, 30] interval. |
| LNDEV4 | The probability that the cell's land evaporation rate is within the [30, 40] interval. |
| LNDEV5 | The probability that the cell's land evaporation rate is within the [40, 50] interval. |
| LNDEV6 | The probability that the cell's land evaporation rate is within the [50, 60] interval. |
| LNDEV7 | The probability that the cell's land evaporation rate is within the [60, 70] interval. |
| LNDEV8 | The probability that the cell's land evaporation rate is within the [70, 80] interval. |
| LNDEV9 | The probability that the cell's land evaporation rate is within the [80, 90] interval. |
| LNDEV10 | The probability that the cell's land evaporation rate is within the [90, 100] interval. |
| WS10M_MAX1 | The probability that the cell's maximum wind speed at 10 meters is within the [0, 10] interval. |
| WS10M_MAX2 | The probability that the cell's maximum wind speed at 10 meters is within the [10, 20] interval. |
| WS10M_MAX3 | The probability that the cell's maximum wind speed at 10 meters is within the [20, 30] interval. |
| WS10M_MAX4 | The probability that the cell's maximum wind speed at 10 meters is within the [30, 40] interval. |
| WS10M_MAX5 | The probability that the cell's maximum wind speed at 10 meters is within the [40, 50] interval. |
| WS10M_MAX6 | The probability that the cell's maximum wind speed at 10 meters is within the [50, 60] interval. |
| WS10M_MAX7 | The probability that the cell's maximum wind speed at 10 meters is within the [60, 70] interval. |
| WS10M_MAX8 | The probability that the cell's maximum wind speed at 10 meters is within the [70, 80] interval. |
| WS10M_MAX9 | The probability that the cell's maximum wind speed at 10 meters is within the [80, 90] interval. |
| WS10M_MAX10 | The probability that the cell's maximum wind speed at 10 meters is within the [90, 100] interval. |
| WS10M_MIN1 | The probability that the cell's minimum wind speed at 10 meters is within the [0, 10] interval. |
| WS10M_MIN2 | The probability that the cell's minimum wind speed at 10 meters is within the [10, 20] interval. |
| WS10M_MIN3 | The probability that the cell's minimum wind speed at 10 meters is within the [20, 30] interval. |
| WS10M_MIN4 | The probability that the cell's minimum wind speed at 10 meters is within the [30, 40] interval. |
| WS10M_MIN5 | The probability that the cell's minimum wind speed at 10 meters is within the [40, 50] interval. |
| WS10M_MIN6 | The probability that the cell's minimum wind speed at 10 meters is within the [50, 60] interval. |
| WS10M_MIN7 | The probability that the cell's minimum wind speed at 10 meters is within the [60, 70] interval. |
| WS10M_MIN8 | The probability that the cell's minimum wind speed at 10 meters is within the [70, 80] interval. |
| WS10M_MIN9 | The probability that the cell's minimum wind speed at 10 meters is within the [80, 90] interval. |
| WS10M_MIN10 | The probability that the cell's minimum wind speed at 10 meters is within the [90, 100] interval. |
| RH2M1 | The probability that the cell's relative humidity at 2 meters is within the [0, 10] interval. |
| RH2M2 | The probability that the cell's relative humidity at 2 meters is within the [10, 20] interval. |
| RH2M3 | The probability that the cell's relative humidity at 2 meters is within the [20, 30] interval. |
| RH2M4 | The probability that the cell's relative humidity at 2 meters is within the [30, 40] interval. |
| RH2M5 | The probability that the cell's relative humidity at 2 meters is within the [40, 50] interval. |
| RH2M6 | The probability that the cell's relative humidity at 2 meters is within the [50, 60] interval. |
| RH2M7 | The probability that the cell's relative humidity at 2 meters is within the [60, 70] interval. |
| RH2M8 | The probability that the cell's relative humidity at 2 meters is within the [70, 80] interval. |
| RH2M9 | The probability that the cell's relative humidity at 2 meters is within the [80, 90] interval. |
| RH2M10 | The probability that the cell's relative humidity at 2 meters is within the [90, 100] interval. |
| PRECTOTCORR1 | The probability that the cell's corrected total precipitation is within the [0, 10] interval. |
| PRECTOTCORR2 | The probability that the cell's corrected total precipitation is within the [10, 20] interval. |
| PRECTOTCORR3 | The probability that the cell's corrected total precipitation is within the [20, 30] interval. |
| PRECTOTCORR4 | The probability that the cell's corrected total precipitation is within the [30, 40] interval. |
| PRECTOTCORR5 | The probability that the cell's corrected total precipitation is within the [40, 50] interval. |
| PRECTOTCORR6 | The probability that the cell's corrected total precipitation is within the [50, 60] interval. |
| PRECTOTCORR7 | The probability that the cell's corrected total precipitation is within the [60, 70] interval. |
| PRECTOTCORR8 | The probability that the cell's corrected total precipitation is within the [70, 80] interval. |
| PRECTOTCORR9 | The probability that the cell's corrected total precipitation is within the [80, 90] interval. |
| PRECTOTCORR10 | The probability that the cell's corrected total precipitation is within the [90, 100] interval. |
| T2M_MAX1 | The probability that the cell's maximum temperature at 2 meters is within the [0, 10] interval. |
| T2M_MAX2 | The probability that the cell's maximum temperature at 2 meters is within the [10, 20] interval. |
| T2M_MAX3 | The probability that the cell's maximum temperature at 2 meters is within the [20, 30] interval. |
| T2M_MAX4 | The probability that the cell's maximum temperature at 2 meters is within the [30, 40] interval. |
| T2M_MAX5 | The probability that the cell's maximum temperature at 2 meters is within the [40, 50] interval. |
| T2M_MAX6 | The probability that the cell's maximum temperature at 2 meters is within the [50, 60] interval. |
| T2M_MAX7 | The probability that the cell's maximum temperature at 2 meters is within the [60, 70] interval. |
| T2M_MAX8 | The probability that the cell's maximum temperature at 2 meters is within the [70, 80] interval. |
| T2M_MAX9 | The probability that the cell's maximum temperature at 2 meters is within the [80, 90] interval. |
| T2M_MAX10 | The probability that the cell's maximum temperature at 2 meters is within the [90, 100] interval. |
| T2M_MIN1 | The probability that the cell's minimum temperature at 2 meters is within the [0, 10] interval. |
| T2M_MIN2 | The probability that the cell's minimum temperature at 2 meters is within the [10, 20] interval. |
| T2M_MIN3 | The probability that the cell's minimum temperature at 2 meters is within the [20, 30] interval. |
| T2M_MIN4 | The probability that the cell's minimum temperature at 2 meters is within the [30, 40] interval. |
| T2M_MIN5 | The probability that the cell's minimum temperature at 2 meters is within the [40, 50] interval. |
| T2M_MIN6 | The probability that the cell's minimum temperature at 2 meters is within the [50, 60] interval. |
| T2M_MIN7 | The probability that the cell's minimum temperature at 2 meters is within the [60, 70] interval. |
| T2M_MIN8 | The probability that the cell's minimum temperature at 2 meters is within the [70, 80] interval. |
| T2M_MIN9 | The probability that the cell's minimum temperature at 2 meters is within the [80, 90] interval. |
| T2M_MIN10 | The probability that the cell's minimum temperature at 2 meters is within the [90, 100] interval. |

**Appendix B:** Full Performance Metrics of Machine Learning Classifiers

100 km Grid Cell Granularity

| Classifier | Precision | Recall | F1-Score | AUC |
|---|---|---|---|---|
| Decision Tree conflict class | 0.83 | 1 | 0.91 | 0.99 |
| Random Forest conflict class | 1 | 0.6 | 0.75 | 0.98 |
| AdaBoost conflict class | 0.83 | 1 | 0.91 | 0.99 |
| Linear SVM conflict class | 0.67 | 0.4 | 0.5 | 0.98 |
| Logistic Regression conflict class | 0.67 | 0.8 | 0.73 | 0.96 |
| MLP conflict class | 0.6 | 0.6 | 0.6 | 0.85 |
| NN conflict class | 1 | 0.6 | 0.75 | 0.8 |

**Table 8.** Performance metrics of different classifiers for predicting pastoralist attacks in Cameroon - granularity 100 km.

| Classifier | Precision | Recall | F1-Score | AUC |
|---|---|---|---|---|
| Decision Tree conflict class | 0.73 | 0.57 | 0.64 | 0.75 |
| Random Forest conflict class | 0.83 | 0.36 | 0.50 | 0.92 |
| AdaBoost conflict class | 0.69 | 0.64 | 0.67 | 0.77 |
| Linear SVM conflict class | 1.00 | 0.57 | 0.73 | 0.82 |
| Logistic Regression conflict class | 0.73 | 0.57 | 0.64 | 0.88 |
| MLP conflict class | 0.54 | 0.50 | 0.52 | 0.76 |
| NN conflict class | 0.78 | 0.50 | 0.61 | 0.72 |

**Table 9.** Performance metrics of different classifiers for predicting pastoralist attacks in the Central African Republic - granularity 100 km.

| Classifier | Precision | Recall | F1-Score | AUC |
|---|---|---|---|---|
| Decision Tree conflict class | 0.50 | 0.25 | 0.33 | 0.60 |
| Random Forest conflict class | 0.33 | 0.12 | 0.18 | 0.91 |
| AdaBoost conflict class | 0.50 | 0.38 | 0.43 | 0.66 |
| Linear SVM conflict class | 0.40 | 0.50 | 0.44 | 0.83 |
| Logistic Regression conflict class | 0.50 | 0.25 | 0.33 | 0.91 |
| MLP conflict class | 0.11 | 0.12 | 0.12 | 0.74 |
| NN conflict class | 0.43 | 0.38 | 0.40 | 0.65 |

**Table 10.** Performance metrics of different classifiers for predicting pastoralist attacks in Chad - granularity 100 km.

| Classifier | Precision | Recall | F1-Score | AUC |
|---|---|---|---|---|
| Decision Tree conflict class | 0.50 | 0.33 | 0.40 | 0.89 |
| Random Forest conflict class | 0.75 | 0.50 | 0.60 | 0.98 |
| AdaBoost conflict class | 0.71 | 0.83 | 0.77 | 0.98 |
| Linear SVM conflict class | 0.67 | 0.67 | 0.67 | 0.98 |
| Logistic Regression conflict class | 1.00 | 0.17 | 0.29 | 0.98 |
| MLP conflict class | 0.30 | 0.50 | 0.37 | 0.94 |
| NN conflict class | 0.44 | 0.67 | 0.53 | 0.82 |

**Table 11.** Performance metrics of different classifiers for predicting pastoralist attacks in the Democratic Republic of Congo - granularity 100 km.

75 km Grid Cell Granularity

| Classifier | Precision | Recall | F1-Score | AUC |
| --- | --- | --- | --- | --- |
| Decision Tree conflict class | 0.29 | 0.33 | 0.31 | 0.62 |
| Random Forest conflict class | 0.00 | 0.00 | 0.00 | 0.78 |
| AdaBoost conflict class | 0.33 | 0.17 | 0.22 | 0.57 |
| Linear SVM conflict class | 1.00 | 0.17 | 0.29 | 0.84 |
| Logistic Regression conflict class | 1.00 | 0.17 | 0.29 | 0.79 |
| MLP conflict class | 0.50 | 0.50 | 0.50 | 0.83 |
| NN conflict class | 1.00 | 0.33 | 0.50 | 0.67 |

**Table 12.** Performance metrics of different classifiers for predicting pastoralist attacks in Cameroon - granularity 75 km.

| Classifier | Precision | Recall | F1-Score | AUC |
| --- | --- | --- | --- | --- |
| Decision Tree conflict class | 0.50 | 0.31 | 0.38 | 0.79 |
| Random Forest conflict class | 0.86 | 0.46 | 0.60 | 0.94 |
| AdaBoost conflict class | 0.43 | 0.46 | 0.44 | 0.76 |
| Linear SVM conflict class | 0.67 | 0.46 | 0.55 | 0.92 |
| Logistic Regression conflict class | 0.75 | 0.69 | 0.72 | 0.95 |
| MLP conflict class | 0.42 | 0.77 | 0.54 | 0.77 |
| NN conflict class | 0.67 | 0.77 | 0.71 | 0.85 |

**Table 13.** Performance metrics of different classifiers for predicting pastoralist attacks in the Central African Republic - granularity 75 km.

| Classifier | Precision | Recall | F1-Score | AUC |
|---|---|---|---|---|
| Decision Tree conflict class | 0.33 | 0.22 | 0.27 | 0.68 |
| Random Forest conflict class | 1.00 | 0.11 | 0.20 | 0.87 |
| AdaBoost conflict class | 0.50 | 0.33 | 0.40 | 0.89 |
| Linear SVM conflict class | 0.71 | 0.56 | 0.63 | 0.88 |
| Logistic Regression conflict class | 0.50 | 0.22 | 0.31 | 0.86 |
| MLP conflict class | 0.50 | 0.33 | 0.40 | 0.85 |
| NN conflict class | 0.67 | 0.44 | 0.53 | 0.71 |

**Table 14.** Performance metrics of different classifiers for predicting pastoralist attacks in Chad - granularity 75 km.

| Classifier | Precision | Recall | F1-Score | AUC |
|---|---|---|---|---|
| Decision Tree conflict class | 0.41 | 0.64 | 0.50 | 0.63 |
| Random Forest conflict class | 0.40 | 0.18 | 0.25 | 0.98 |
| AdaBoost conflict class | 0.47 | 0.64 | 0.54 | 0.90 |
| Linear SVM conflict class | 0.40 | 0.18 | 0.25 | 0.97 |
| Logistic Regression conflict class | 0.50 | 0.36 | 0.42 | 0.98 |
| MLP conflict class | 0.60 | 0.55 | 0.57 | 0.90 |
| NN conflict class | 0.71 | 0.46 | 0.56 | 0.72 |

**Table 15.** Performance metrics of different classifiers for predicting pastoralist attacks in the Democratic Republic of Congo - granularity 75 km.

50 km Grid Cell Granularity

| Classifier | Precision | Recall | F1-Score | AUC |
|---|---|---|---|---|
| Decision Tree conflict class | 0.50 | 0.60 | 0.55 | 0.79 |
| Random Forest conflict class | 0.50 | 0.20 | 0.29 | 0.99 |
| AdaBoost conflict class | 0.50 | 0.40 | 0.44 | 0.87 |
| Linear SVM conflict class | 1.00 | 0.20 | 0.33 | 0.98 |
| Logistic Regression conflict class | 0.25 | 0.20 | 0.22 | 0.97 |
| MLP conflict class | 0.40 | 0.40 | 0.40 | 0.84 |
| NN conflict class | 0.57 | 0.80 | 0.67 | 0.89 |

**Table 16.** Performance metrics of different classifiers for predicting pastoralist attacks in Cameroon - granularity 50 km.

| Classifier | Precision | Recall | F1-Score | AUC |
|---|---|---|---|---|
| Decision Tree conflict class | 0.55 | 0.32 | 0.40 | 0.73 |
| Random Forest conflict class | 0.50 | 0.16 | 0.24 | 0.87 |
| AdaBoost conflict class | 0.25 | 0.16 | 0.19 | 0.69 |
| Linear SVM conflict class | 0.60 | 0.16 | 0.25 | 0.86 |
| Logistic Regression conflict class | 0.40 | 0.11 | 0.17 | 0.88 |
| MLP conflict class | 0.24 | 0.21 | 0.22 | 0.72 |
| NN conflict class | 0.56 | 0.26 | 0.36 | 0.62 |

**Table 17.** Performance metrics of different classifiers for predicting pastoralist attacks in the Central African Republic - granularity 50 km.

| Classifier | Precision | Recall | F1-Score | AUC |
| --- | --- | --- | --- | --- |
| Decision Tree conflict class | 0.20 | 0.08 | 0.11 | 0.68 |
| Random Forest conflict class | 1.00 | 0.15 | 0.27 | 0.88 |
| AdaBoost conflict class | 0.33 | 0.15 | 0.21 | 0.79 |
| Linear SVM conflict class | 0.24 | 0.31 | 0.27 | 0.85 |
| Logistic Regression conflict class | 0.75 | 0.23 | 0.35 | 0.88 |
| MLP conflict class | 0.27 | 0.31 | 0.29 | 0.75 |
| NN conflict class | 0.44 | 0.31 | 0.36 | 0.64 |

**Table 18.** Performance metrics of different classifiers for predicting pastoralist attacks in Chad - granularity 50 km.

| Classifier | Precision | Recall | F1-Score | AUC |
| --- | --- | --- | --- | --- |
| Decision Tree conflict class | 0.50 | 0.50 | 0.50 | 0.92 |
| Random Forest conflict class | 0.50 | 0.11 | 0.18 | 0.94 |
| AdaBoost conflict class | 0.53 | 0.56 | 0.54 | 0.89 |
| Linear SVM conflict class | 0.50 | 0.17 | 0.25 | 0.91 |
| Logistic Regression conflict class | 0.78 | 0.39 | 0.52 | 0.89 |
| MLP conflict class | 0.46 | 0.33 | 0.39 | 0.95 |
| NN conflict class | 0.44 | 0.44 | 0.44 | 0.71 |

**Table 19.** Performance metrics of different classifiers for predicting pastoralist attacks in the Democratic Republic of Congo - granularity 50 km.

# Appendix C: Supplementary Tables for Univariate Hypotheses Testing

| Variable | Difference in mean | 95% CI, lower | 95% CI, upper | Bonferroni P-value |
|----------|--------------------|---------------|---------------|--------------------|
| LAI6     | -0.0472            | -0.0679       | -0.0266       | 0.002              |
| LAI7     | -0.0388            | -0.0544       | -0.0232       | 0.0003             |
| LAI8     | -0.0349            | -0.0504       | -0.0194       | 0.0018             |
| LAI9     | -0.029             | -0.0417       | -0.0162       | 0.0017             |
| LAI10    | -0.0279            | -0.0404       | -0.0154       | 0.0023             |
| LNDEV10  | -0.0068            | -0.0099       | -0.0037       | 0.0038             |

**Table 20.** Cameroon -Results from the Statistical Analysis of Univariate Meteorological and Terrain-Related Feature Hypotheses.

| Variable | Difference in mean | 95% CI, lower | 95% CI, upper | Bonferroni P-value |
|---|---|---|---|---|
| LAI2 | 7.34E-02 | 3.79E-02 | 1.09E-01 | 1.15E-02 |
| LAI3 | 4.22E-02 | 2.41E-02 | 6.03E-02 | 1.11E-03 |
| LAI4 | 8.91E-02 | 5.05E-02 | 1.28E-01 | 2.67E-03 |
| LAI5 | 1.12E-01 | 6.15E-02 | 1.63E-01 | 5.64E-03 |
| LAI8 | -5.39E-02 | -7.17E-02 | -3.61E-02 | 1.93E-06 |
| LAI9 | -4.70E-02 | -6.58E-02 | -2.81E-02 | 2.54E-04 |
| LAI10 | -5.17E-02 | -7.49E-02 | -2.85E-02 | 2.34E-03 |
| GRN6 | -7.85E-02 | -1.13E-01 | -4.41E-02 | 1.42E-03 |
| SSW1 | -5.43E-02 | -7.73E-02 | -3.12E-02 | 8.01E-04 |
| SSW2 | -5.38E-02 | -7.11E-02 | -3.65E-02 | 7.87E-07 |
| SST9 | -3.64E-02 | -4.95E-02 | -2.34E-02 | 1.65E-05 |
| SST10 | -9.69E-03 | -1.38E-02 | -5.59E-03 | 7.63E-04 |
| LNDEV1 | -1.06E-01 | -1.41E-01 | -7.17E-02 | 1.22E-06 |
| LNDEV7 | 1.19E-01 | 7.91E-02 | 1.59E-01 | 1.54E-05 |
| T2M_MAX9 | -3.22E-02 | -4.64E-02 | -1.80E-02 | 1.96E-03 |
| T2M_MIN8 | -3.44E-02 | -4.74E-02 | -2.15E-02 | 6.66E-05 |
| T2M_MIN9 | -1.35E-02 | -2.02E-02 | -6.84E-03 | 1.23E-02 |
| RH2M1 | -4.34E-02 | -5.96E-02 | -2.71E-02 | 5.64E-05 |
| RH2M2 | -4.85E-02 | -6.69E-02 | -3.00E-02 | 9.76E-05 |
| RH2M7 | 3.42E-02 | 2.02E-02 | 4.82E-02 | 9.13E-04 |
| RH2M8 | 6.85E-02 | 4.54E-02 | 9.15E-02 | 2.18E-05 |
| RH2M9 | 8.82E-02 | 4.71E-02 | 1.29E-01 | 6.38E-03 |
| RH2M10 | -1.09E-01 | -1.42E-01 | -7.52E-02 | 1.75E-07 |
| WS10M_MIN1 | -1.58E-01 | -2.22E-01 | -9.44E-02 | 2.53E-04 |
| WS10M_MIN2 | 1.05E-01 | 7.68E-02 | 1.34E-01 | 2.61E-09 |
| WS10M_MIN3 | 8.92E-02 | 5.89E-02 | 1.20E-01 | 8.65E-06 |
| WS10M_MIN7 | -9.00E-03 | -1.23E-02 | -5.65E-03 | 3.94E-05 |
| WS10M_MIN8 | -2.78E-03 | -4.19E-03 | -1.38E-03 | 1.50E-02 |
| WS10M_MAX1 | -2.41E-01 | -3.17E-01 | -1.65E-01 | 3.37E-07 |
| WS10M_MAX2 | 1.29E-01 | 8.48E-02 | 1.73E-01 | 2.58E-05 |
| WS10M_MAX3 | 1.39E-01 | 1.02E-01 | 1.76E-01 | 7.22E-09 |
| WS10M_MAX6 | -3.01E-02 | -4.45E-02 | -1.58E-02 | 6.63E-03 |
| WS10M_MAX7 | -2.05E-02 | -2.97E-02 | -1.13E-02 | 2.65E-03 |

**Table 21.** CAR - Results from the Statistical Analysis of Univariate Meteorological and Terrain-Related Feature Hypotheses.

| Variable | Difference in mean | 95% CI, lower | 95% CI, upper | Bonferroni P-value |
|---|---|---|---|---|
| LAI1 | -3.71E-01 | -4.96E-01 | -2.46E-01 | 5.19E-05 |
| LAI2 | 1.78E-01 | 1.38E-01 | 2.19E-01 | 5.89E-09 |
| LAI3 | 7.10E-02 | 4.41E-02 | 9.78E-02 | 4.42E-04 |
| GRN1 | -2.16E-02 | -3.24E-02 | -1.09E-02 | 1.14E-02 |
| GRN2 | -1.26E-02 | -1.88E-02 | -6.39E-03 | 9.53E-03 |
| GRN3 | -2.66E-02 | -3.67E-02 | -1.65E-02 | 5.71E-05 |
| GRN4 | -4.50E-02 | -5.90E-02 | -3.09E-02 | 2.12E-07 |
| GRN6 | -8.58E-02 | -1.08E-01 | -6.34E-02 | 2.97E-10 |
| GRN7 | -1.23E-01 | -1.64E-01 | -8.17E-02 | 8.40E-06 |
| GRN9 | 2.51E-01 | 1.89E-01 | 3.14E-01 | 5.12E-08 |
| GRN10 | 1.66E-01 | 9.13E-02 | 2.41E-01 | 6.32E-03 |
| SSW4 | 4.73E-02 | 2.90E-02 | 6.55E-02 | 6.30E-04 |
| SSW5 | 4.24E-02 | 2.88E-02 | 5.60E-02 | 8.82E-06 |
| SSW6 | 4.64E-02 | 2.90E-02 | 6.39E-02 | 2.92E-04 |
| SSW7 | 5.74E-02 | 4.06E-02 | 7.42E-02 | 1.01E-06 |
| SSW8 | 6.48E-02 | 3.48E-02 | 9.49E-02 | 1.18E-02 |
| SST1 | -1.31E-02 | -1.72E-02 | -9.08E-03 | 1.39E-07 |
| SST5 | -6.02E-02 | -9.08E-02 | -2.95E-02 | 2.64E-02 |
| SST6 | 1.82E-01 | 1.34E-01 | 2.30E-01 | 2.23E-07 |
| SST7 | 6.88E-02 | 3.87E-02 | 9.89E-02 | 5.87E-03 |
| LNDEV1 | -3.08E-01 | -3.91E-01 | -2.26E-01 | 3.11E-08 |
| LNDEV3 | 3.36E-02 | 2.05E-02 | 4.67E-02 | 6.62E-04 |
| LNDEV4 | 3.53E-02 | 2.50E-02 | 4.56E-02 | 1.14E-06 |
| LNDEV5 | 3.69E-02 | 2.49E-02 | 4.89E-02 | 1.63E-05 |
| LNDEV6 | 5.82E-02 | 4.12E-02 | 7.53E-02 | 3.39E-06 |
| LNDEV7 | 1.03E-01 | 6.51E-02 | 1.41E-01 | 2.89E-04 |
| T2M_MAX1 | -4.99E-03 | -6.80E-03 | -3.17E-03 | 1.94E-05 |
| T2M_MAX2 | -1.60E-02 | -2.01E-02 | -1.18E-02 | 1.82E-10 |
| T2M_MAX6 | 5.32E-02 | 3.77E-02 | 6.88E-02 | 1.71E-07 |
| T2M_MIN2 | -1.39E-02 | -1.75E-02 | -1.02E-02 | 2.73E-10 |
| T2M_MIN5 | -4.70E-02 | -6.11E-02 | -3.29E-02 | 2.52E-06 |
| T2M_MIN8 | 1.60E-01 | 1.18E-01 | 2.01E-01 | 1.88E-07 |
| RH2M2 | -1.01E-01 | -1.27E-01 | -7.53E-02 | 9.55E-10 |
| RH2M3 | -6.95E-02 | -8.75E-02 | -5.15E-02 | 1.75E-08 |
| RH2M7 | 3.13E-02 | 1.97E-02 | 4.29E-02 | 2.46E-04 |
| RH2M8 | 4.48E-02 | 3.24E-02 | 5.73E-02 | 1.43E-07 |
| RH2M9 | 9.52E-02 | 6.71E-02 | 1.23E-01 | 1.77E-06 |
| RH2M10 | 3.85E-02 | 2.12E-02 | 5.59E-02 | 6.33E-03 |
| PRECTOTCORR1 | -3.39E-02 | -4.58E-02 | -2.19E-02 | 1.86E-04 |
| PRECTOTCORR2 | 2.38E-02 | 1.49E-02 | 3.27E-02 | 4.78E-04 |
| WS10M_MIN1 | 1.06E-01 | 6.76E-02 | 1.44E-01 | 1.45E-04 |
| WS10M_MIN2 | 8.02E-02 | 5.38E-02 | 1.06E-01 | 1.85E-05 |
| WS10M_MIN4 | -1.04E-01 | -1.36E-01 | -7.25E-02 | 2.21E-06 |
| WS10M_MIN5 | -3.17E-02 | -4.42E-02 | -1.92E-02 | 3.78E-04 |
| WS10M_MIN7 | -4.20E-03 | -6.24E-03 | -2.16E-03 | 7.53E-03 |
| WS10M_MIN8 | 3.83E-04 | -1.97E-03 | -5.79E-04 | 4.22E-02 |
| WS10M_MAX2 | 9.93E-02 | 6.81E-02 | 1.30E-01 | 5.61E-06 |
| WS10M_MAX3 | 5.30E-02 | 2.70E-02 | 7.89E-02 | 1.93E-02 |
| WS10M_MAX4 | -4.90E-02 | -7.18E-02 | -2.62E-02 | 7.37E-03 |
| WS10M_MAX5 | -6.19E-02 | -9.25E-02 | -3.14E-02 | 2.25E-02 |
| WS10M_MAX6 | -4.38E-02 | -6.25E-02 | -2.51E-02 | 2.19E-03 |

**Table 22.** Chad - Results from the Statistical Analysis of Univariate Meteorological and Terrain-Related Feature Hypotheses.

| Variable | Difference in mean | 95% CI, lower | 95% CI, upper | Bonferroni P-value |
|---|---|---|---|---|
| LAI5 | -6.42E-02 | -8.69E-02 | -4.15E-02 | 2.21E-04 |
| LAI8 | 1.34E-01 | 8.78E-02 | 1.81E-01 | 1.65E-04 |
| LAI9 | 1.27E-01 | 7.38E-02 | 1.80E-01 | 3.35E-03 |
| GRN3 | -3.79E-02 | -4.64E-02 | -2.93E-02 | 4.26E-10 |
| GRN4 | -3.01E-02 | -4.31E-02 | -1.71E-02 | 4.07E-03 |
| GRN7 | 1.00E-01 | 5.89E-02 | 1.41E-01 | 2.59E-03 |
| SST1 | -7.00E-04 | -1.01E-03 | -3.89E-04 | 1.32E-03 |
| SST6 | -1.62E-01 | -2.16E-01 | -1.08E-01 | 5.61E-05 |
| SST7 | -6.45E-02 | -8.30E-02 | -4.59E-02 | 1.11E-06 |
| SST8 | -1.92E-02 | -2.67E-02 | -1.17E-02 | 5.40E-04 |
| LNDEV8 | -6.04E-02 | -8.88E-02 | -3.20E-02 | 1.56E-02 |
| T2M_MAX8 | -9.16E-02 | -1.36E-01 | -4.69E-02 | 2.17E-02 |
| T2M_MIN2 | -3.16E-04 | -4.92E-04 | -1.41E-04 | 4.80E-02 |
| T2M_MIN3 | -1.04E-02 | -1.33E-02 | -7.54E-03 | 8.83E-10 |
| T2M_MIN4 | -2.41E-02 | -3.65E-02 | -1.18E-02 | 3.00E-02 |
| T2M_MIN8 | -1.95E-01 | -2.73E-01 | -1.17E-01 | 1.00E-03 |
| RH2M2 | -2.33E-03 | -3.20E-03 | -1.46E-03 | 2.41E-05 |
| RH2M4 | -2.87E-02 | -4.13E-02 | -1.61E-02 | 4.69E-03 |
| RH2M8 | 8.99E-02 | 5.26E-02 | 1.27E-01 | 2.75E-03 |
| WS10M_MIN2 | 1.14E-01 | 6.98E-02 | 1.58E-01 | 7.58E-04 |
| WS10M_MIN5 | -3.45E-02 | -4.87E-02 | -2.04E-02 | 1.27E-03 |
| WS10M_MIN6 | -1.07E-02 | -1.39E-02 | -7.40E-03 | 2.99E-07 |
| WS10M_MIN7 | -3.34E-03 | -4.28E-03 | -2.40E-03 | 1.06E-09 |
| WS10M_MIN8 | -1.11E-03 | -1.55E-03 | -6.63E-04 | 1.48E-04 |
| WS10M_MAX1 | -1.39E-01 | -1.80E-01 | -9.80E-02 | 6.15E-07 |
| WS10M_MAX3 | 1.79E-01 | 1.24E-01 | 2.35E-01 | 1.74E-05 |
| WS10M_MAX7 | -1.68E-02 | -2.37E-02 | -9.83E-03 | 1.12E-03 |
| WS10M_MAX8 | -5.53E-03 | -8.32E-03 | -2.74E-03 | 2.34E-02 |

**Table 23.** DRC - Results from the Statistical Analysis of Univariate Meteorological and Terrain-Related Feature Hypotheses.

**Appendix D:** Respective Decision Trees for Each Univariate Hypotheses

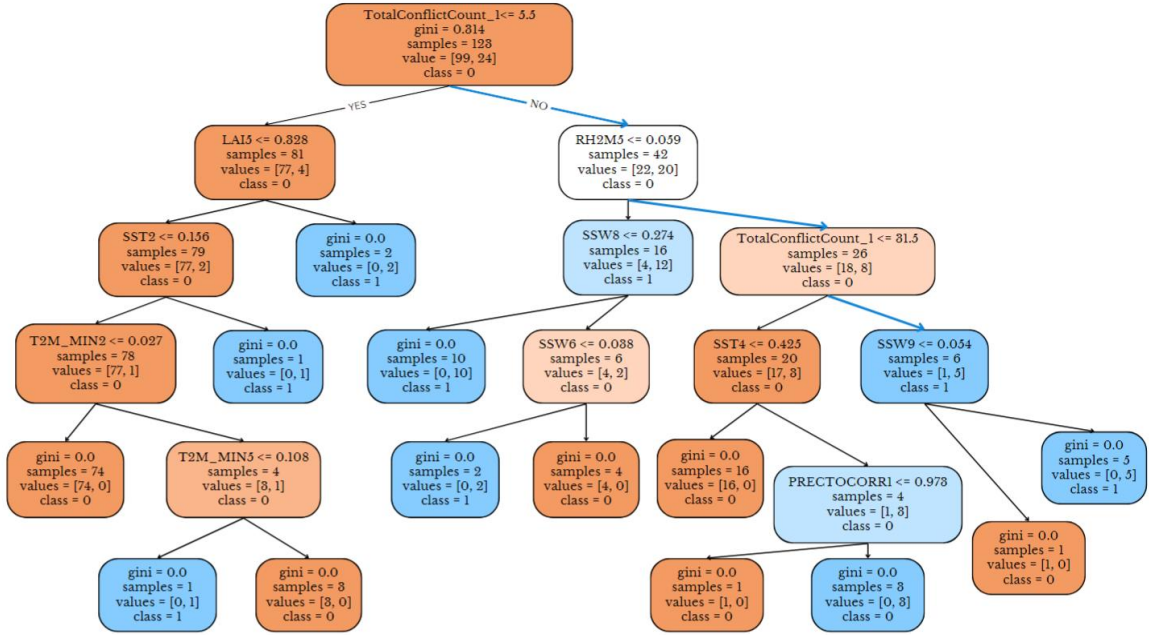

**Figure 10.** Highlighted Path in Learned Decision Tree for CAR 100 km granularity - hypothesis 5.

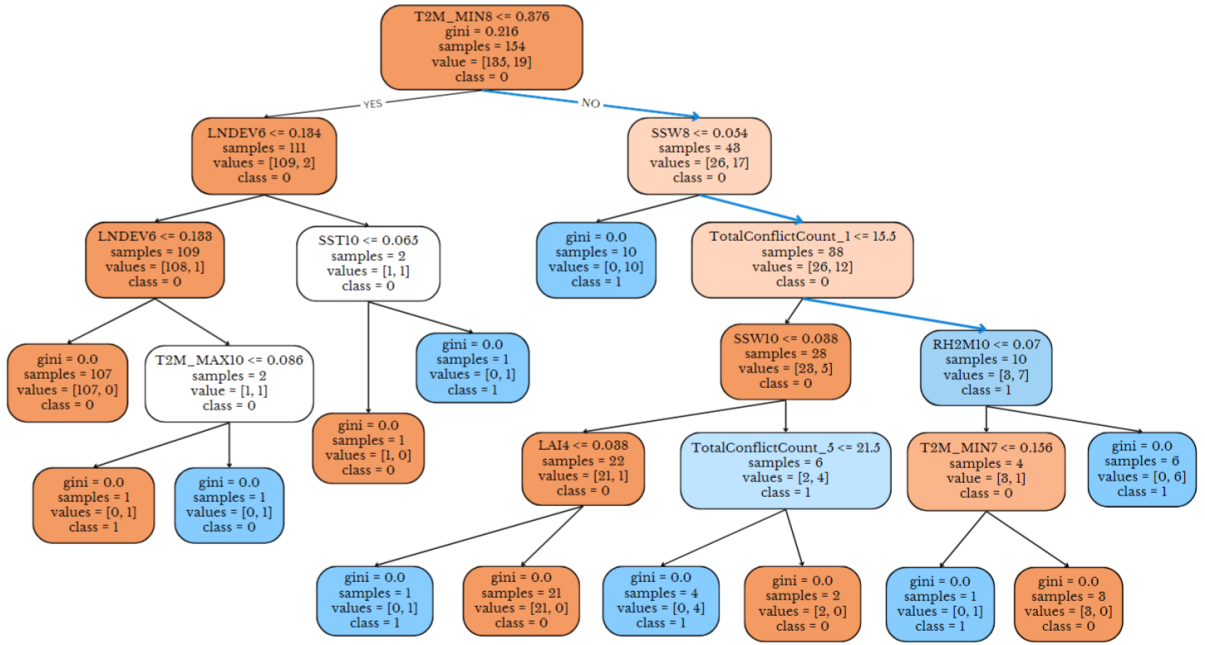

**Figure 11.** Highlighted Path in Learned Decision Tree for Chad 100 km granularity - hypothesis 7.

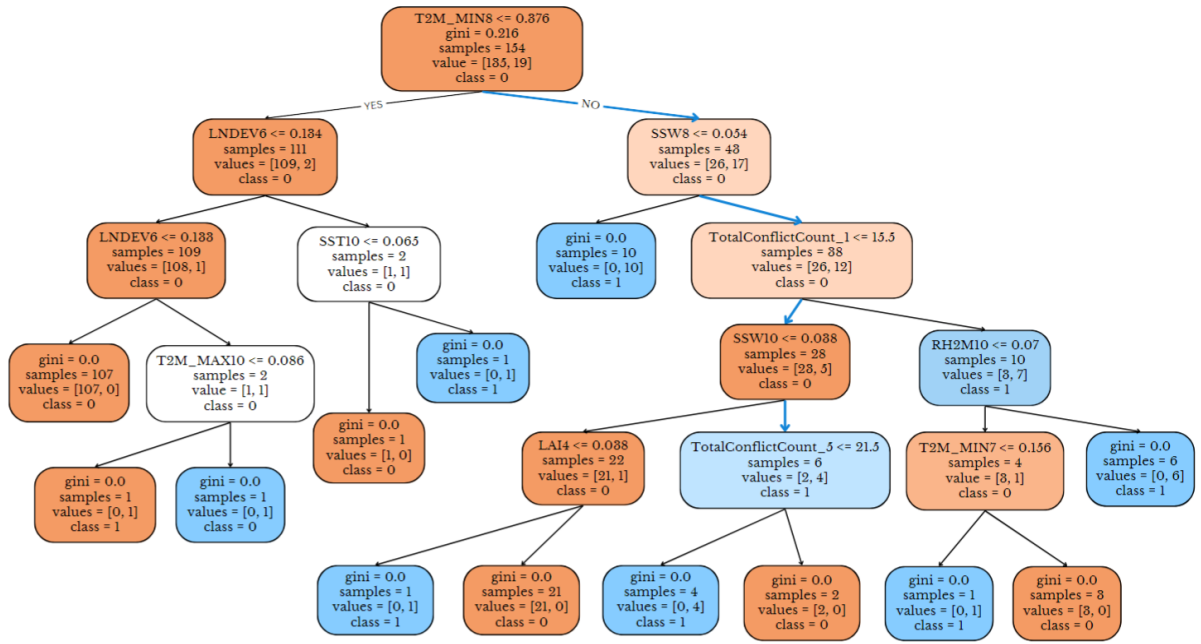

**Figure 12.** Highlighted Path in Learned Decision Tree for Chad 100 km granularity – hypothesis 8.

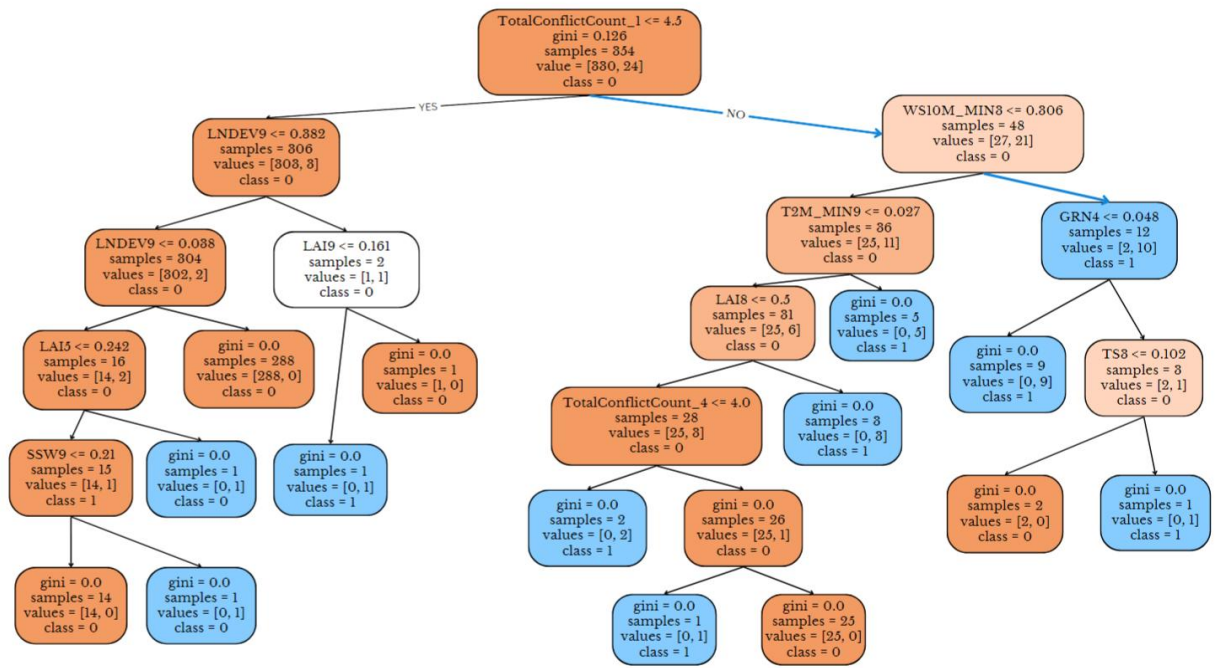

**Figure 13.** Highlighted Path in Learned Decision Tree for DRC 100 km granularity – hypothesis 10.